\documentclass[twocolumn,showpacs,amsmath,amssymb,superscriptaddress,aps,prc]{revtex4-1}
\usepackage[dvips]{graphicx}
\usepackage{epsfig}
\usepackage{amsmath}
\usepackage[dvips]{color}
\newcommand{\beq}{\begin{eqnarray}}
\newcommand{\eeq}{\end{eqnarray}}

\newcommand{\li}{\mbox{$^7$Li}}
\newcommand{\at}{\mbox{$\alpha+t$}}
\newcommand{\lipb}{\mbox{$^7$Li+$^{208}$Pb}}
\newcommand{\emax}{\mbox{$E_{\rm max}$}}
\newcommand{\lmax}{\mbox{$\ell_{\rm max}$}}

\begin{document}
	\title{Coulomb and nuclear effects in breakup and reaction cross sections}
	\author{P. Descouvemont}
	\email{pdesc@ulb.ac.be}
\affiliation{Physique Nucl\'eaire Th\'eorique et Physique Math\'ematique, C.P. 229,
	Universit\'e Libre de Bruxelles (ULB), B 1050 Brussels, Belgium}
	\author{L. F. Canto}
	\email{canto@if.ufrj.br}
	\affiliation{Instituto de F\'isica, Universidade Federal do Rio de Janeiro, C.P. 68528, 21941-972 Rio de Janeiro, RJ, Brazil}
	\affiliation{Instituto de Física, Universidade Federal Fluminense, Av. Gal. Milton Tavares de Souza s/n, Niter\'{o}i, RJ, Brazil}
	\author{M. S. Hussein}
	\email{hussein@if.usp.br}
	\affiliation{Departamento de F\'isica Matem\'atica, Instituto de F\'isica, Universidade de S\~ao Paulo, C.P. 66318, 05314-970, S\~ao Paulo, SP, Brazil}
	\affiliation{Instituto de Estudos Avan\c{c}ados, Universidade de S\~ao Paulo, C.P. 72012, 05508-970, 
		S\~ao Paulo, SP, Brazil}
	\affiliation{Departamento de F\'{i}sica, Instituto Tecnol\'{o}gico de Aeron\'{a}utica, CTA, S\~{a}o Jos\'{e} dos Campos, S\~ao Paulo, SP, Brazil}
		\date{\today}
	\begin{abstract}
		We use a three-body Continuum Discretized Coupled Channel (CDCC) model to investigate Coulomb and nuclear
		effects in breakup and reaction cross sections. The breakup of the projectile is simulated by
		a finite number of square integrable wave functions. First we show that the scattering matrices can be
		split in a nuclear term, and in a Coulomb term. This decomposition is based on the
		Lippmann-Schwinger equation, and requires the scattering wave functions. We present two different
		methods to separate both effects.
		Then, we apply this separation to breakup and reaction cross sections of $\lipb$. For breakup, we 
		investigate various aspects, such as the role of the $\at$ continuum, the angular-momentum distribution,
		 and the balance between
		 Coulomb and nuclear effects.
		We show that there is a large ambiguity in defining the 'Coulomb' and 'nuclear' breakup cross sections, since
		both techniques, although providing the same total breakup cross sections, strongly differ for the
		individual components. We suggest a third method which could be efficiently used to address convergence problems
		at large angular momentum. For reaction cross sections, interference effects are smaller, and the nuclear contribution is dominant above the Coulomb barrier. We also draw attention on different definitions of the reaction cross section which exist in the literature, and which may induce small, but significant, differences in the numerical values.
	\end{abstract}
	\maketitle
\section{Introduction}
\label{sec1}

The development of radioactive facilities has opened many new perspectives in the physics of exotic 
nuclei \cite{BHM01,TSK13}. Radioactive beams are now available with high intensities and purities \cite{BNV13,LLG14}. 
Historically, the first experiments were aimed at measuring reaction cross sections \cite{Ta16}, which provide, 
through simple model assumptions, the interaction radius of the projectile.  This technique lead to the 
discovery of halo nuclei, where one or two nucleons are weakly bound to the core, and therefore contribute 
to a radius much larger than expected from the usual $A^{1/3}$ law.

Breakup  is a fundamental process in reactions
involving exotic nuclei, since they present a low separation energy \cite{KYI86}. 
Even the investigation of elastic scattering requires to simulate breakup 
effects to obtain a fair theoretical description of the data.  This requirement is well addressed by the Continuum Discretized Coupled Channel (CDCC) method, introduced in the seventies to analyse deuteron scattering on heavy targets \cite{Ra74b,SYK86}.  In the CDCC method, the breakup of the projectile is simulated by approximate 
continuum states \cite{AIK87,YOM12}.  The couplings of the elastic channel to these continuum channels modify the scattering matrices, and therefore the elastic cross sections.

The theoretical description of elastic scattering is relatively well understood, even for two-neutron halo 
nuclei \cite{MHO04,FCA15}.  The competing breakup process, however, is more complicated, and has been
investigated by different authors (see, for example, Refs.\ \cite{DLV96,NT99,TNT01,GBC06,MCD08,CDM11,KM12,RLC16}).  
In particular, the interplay between Coulomb and nuclear effects in breakup reactions is of great interest \cite{DLV99,HLN06,KB12}.
In the CDCC approach, the breakup cross section involves transitions to pseudostates (or to bins). The cross section is therefore
a multiple sum over several quantum numbers: total spin
and parity of the system, energy and angular momentum of the projectile. 
These specific breakup problems are on top of the traditional numerical aspects of CDCC (convergence
against the space truncation parameters).
The description of the breakup process is therefore a delicate problem, which deserves careful analyses. 

The main goal of the present work is to address $\lipb$ breakup and reaction cross sections in the CDCC framework. We essentially 
focus on the separation between the nuclear and Coulomb contributions. Knowing 
whether a breakup cross section is dominated by Coulomb effects or not is an important
issue to analyse experiments involving exotic nuclei. This problem has been investigated in the past in simpler models
(see, for example, Refs.\ \cite{DLV98,KB12}), but CDCC approaches are treated in an approximate way \cite{HLN06,CGD15,RLC16}. 
In contrast with the total breakup cross sections, the individual nuclear and Coulomb contributions (as well 
as interferences terms) require the availability of the wave functions.   
A usual approximation consists in determining the individual contributions by neglecting, either Coulomb couplings, or nuclear 
couplings.  
This approach, however, implicitly assumes that the coupling between the elastic and breakup channels is weak, so that
a DWBA approximation is valid. 

The paper is organized as follows.  In Sec.\ \ref{sec2}, we briefly present the main properties of the CDCC model.  Section \ref{sec3} addresses the separation of nuclear and Coulomb components in the scattering matrices and in the cross sections.  The application to $\lipb$ breakup is presented in Secs.\ \ref{sec4} and \ref{sec5}.  We not only discuss the nuclear and Coulomb contributions, but also other important aspects, such as the angular-momentum distribution, the contribution of the different $\alpha + t$ continuum channels, or the convergence against the CDCC truncation parameters. Conclusion and outlook are discussed in Sec.\ \ref{sec6}.

\section{Outline of the CDCC method}
\label{sec2}

The CDCC method has been presented in many papers and reviews (see, for example Refs.\ \cite{AIK87,YOM12}). 
 Here we present a short outline, and define our notations.

Let us first consider the projectile description.  The Hamiltonian is denoted as $H_0(\pmb{\xi})$, where 
$\pmb{\xi}$ is a general 
notation, involving the internal coordinates.  For example, $\pmb{\xi}$ is the relative distance in a two-body description; 
it can represent two coordinates in a three-body model.  In a microscopic approach, $\pmb{\xi}$ stands for all 
nucleon coordinates \cite{DD12}.  For a partial wave $j$ (parity is implied), the Schr\"{o}dinger equation 
of the projectile reads
\beq
H_0(\pmb{\xi})\Phi^{jm}_{0,n}(\pmb{\xi})=E_{0,n}^j \Phi^{jm}_{0,n}(\pmb{\xi}),
\label{eq1}
\eeq
where $n$ is the excitation level in partial wave $j$.

In the CDCC method, the wave functions $\Phi^{jm}_{0,n}$ are expanded, either over a basis (such a Lagrange \cite{Ba15} or Gaussian \cite{HKK03} functions), or over bins \cite{Th88}. Equation (\ref{eq1}) represents a standard eigenvalue problem, where negative solutions correspond to physical states (in $^7$Li, the $j=3/2^-$ and the $j=1/2^-$ bound states). The positive eigenvalues do not have a physical meaning, but correspond to approximations of the continuum.  

In the projectile+target system with relative coordinate $\pmb{R}$, the total wave function is given by the Schr\"{o}dinger equation
\beq
\biggl(-\frac{\hbar^2}{2\mu}\Delta_R+ H_0(\pmb{\xi}) +V(\pmb{R},\pmb{\xi}) \biggr)
\Psi^{JM\pi}_{\omega}=E \Psi^{JM\pi}_{\omega},
\label{eq1b}
\eeq
where $\omega$ is the entrance channel, $\mu$ is the reduced mass, and $V(\pmb{R},\pmb{\xi})$ represents
the total interaction between the constituents of the projectile and the target.
We define a channel function as
\beq
\varphi^{JM\pi}_{jnL}(\Omega_R,\pmb{\xi})=\bigl[ \Phi^{j}_{0,n}(\pmb{\xi}) \otimes Y_L(\Omega_R)\bigr]^{JM},
\label{eq2}
\eeq
where $J$ and $\pi$ are the total angular momentum and parity, and $L$ the relative orbital momentum.  As usual in CDCC calculations, excitations of the target are neglected.  

With the channel functions (\ref{eq2}) we expand the total wave function of the system as
\beq
\Psi^{JM\pi}_{\omega}(\pmb{R},\pmb{\xi})=\frac{1}{R}\sum_c g^{J\pi}_{\omega,c}(R) \varphi^{JM\pi}_{c}(\Omega_R,\pmb{\xi}),
\label{eq3}
\eeq
where $J$ and $\pi$ are the total spin and parity of the system, and index $c$ stands for $(j,n,L)$. 
The summation is truncated in two ways: $\lmax$ (and therefore $j_{\rm max}$) determines the maximum angular momentum in $\li$, and
$\emax$ is the maximum energy of the pseudostates, which, in practice, limits the number of $n$ values.
Stability tests against these two truncation parameters are required to guaranty the reliability of CDCC
calculations.

At center-of-mass (c.m.) energy $E$, the radial functions $g^{J\pi}_{\omega,c}(R)$ are obtained from 
the standard coupled-channel system 
\beq
&&\bigl(T_L +E_{c}-E \bigr) 
g^{J\pi}_{\omega,c}(R) +\sum_{c'}V^{J\pi}_{c,c'}(R) g^{J\pi}_{\omega,c'}(R)=0,
\label{eq4}
\eeq
where the kinetic-energy operator reads
\beq
T_L= -\frac{\hbar^2}{2\mu}\left( \frac{d^2}{dR^2}-\frac{L(L+1)}{R^2}\right),
\label{eq5}
\eeq
and where the coupling potentials $V^{J\pi}_{c,c'}(R)$ are determined from matrix elements of the 
fragment-target interactions between channel functions (\ref{eq2}). 

 At large distance $R$, the 
radial functions tend to a combination of Coulomb functions as
\beq
 g^{J\pi}_{\omega,c}(R)\rightarrow
v_{c}^{-1/2} \Bigl( I_{c} (k_{c}R)\delta_{c \omega} -  O_{c} (k_{c} R)U^{J\pi}_{\omega, c} \Bigr),
\label{eq6}
\eeq
where $I_c(x)$ and $O_c(x)$ are the incoming and outgoing Coulomb functions \cite{Th10}, $k_c$ is the wave 
number in channel $c$, and $U^{J\pi}_{\omega c}$ is an element of the scattering matrix.  It is used 
to compute the various cross sections.  

In the present work, the scattering problem (\ref{eq4}) is solved with 
the $R$-matrix method (see Refs.\ \cite{DB10,De16a}) where coordinate $R$ is split in two regions, 
separated by the channel radius $a$.  In the external region, $a$ is chosen large enough so that the 
asymptotic behaviour (\ref{eq6}) is valid.  In the internal region, the radial functions $g^{J\pi}_{\omega,c}(R)$ 
are expanded over a Lagrange mesh \cite{Ba15}.  This technique permits a fast and accurate calculation of the 
scattering matrix.  Notice that neither the channel radius $a$, nor the number of basis functions $N$, 
should affect the scattering matrices, provided these parameters are properly chosen.  
The stability of the scattering matrices and of the cross sections against variations of $a$ and $N$ is 
a strong constraint on the correctness and precision of the method.  We refer the reader to Refs.\ 
\cite{DB10,De16a,Ba15} for more information.

\section{Coulomb and nuclear amplitudes}
\label{sec3}
\subsection{Scattering matrices}
The basic input to determine the cross sections is the scattering matrix
$\pmb{U}^{J\pi}$. It is computed in the $R$-matrix formalism, as outlined in Sec.\ \ref{sec2}. It is, however,
possible to express the scattering matrix in a fully equivalent way, based on the wave function.
From scattering theory \cite{CH13}, it can be shown that an alternative definition of the scattering
matrix is
\beq
&&U^{J\pi}_{\omega L_{\omega},\alpha L}=\delta_{\omega \alpha}\delta_{L_{\omega}L }+ \nonumber \\
&&\frac{1}{\hbar v_{\alpha}^{1/2}}
\int_0^{\infty} F_{L}(k_{\alpha} R)
\sum_{\alpha' L'} \tilde{V}^{J\pi}_{\alpha L,\alpha' L'}(R)
g^{J\pi}_{\omega L_{\omega},\alpha' L'}(R)dR,
\label{eq10}
\eeq
where we have used normalization (\ref{eq6}) of the wave function, and where
$F_{L}(x)$ is the regular Coulomb function (the Sommerfeld parameter is implied).
Throughout the text, either we use index $c$, or we specifically mention the 
channel $\alpha=j,n$ and the angular momentum $L$ when more appropriate. 
The potentials $\tilde{V}^{J\pi}_{c,c'}(R)$ 
are  defined from
\beq
\tilde{V}^{J\pi}_{c,c'}(R)=V^{J\pi}_{c,c'}(R)-\frac{Z_t Z_p e^2}{R}\delta_{c c'},
\label{eq11}
\eeq
where $Z_t e$ and $Z_p e$ are the charges of the target and of the projectile, respectively.
These potentials contain the nuclear term, and the non-monopole Coulomb
contributions, and are given by
\beq
\tilde{V}^{J\pi}_{c,c'}(R)=\tilde{V}^{J\pi(C)}_{c,c'}(R)+V^{J\pi(N)}_{c,c'}(R).
\label{eq12}
\eeq
The Coulomb terms are known to be important,
as their range can be quite large \cite{DD12b}. 

This method
is referred to as "Method 1", and an element of the scattering matrix 
is decomposed as
\beq
U^{J\pi}_{c,c'}=U^{J\pi(C1)}_{c,c'}+U^{J\pi(N1)}_{c,c'},
\label{eq13}
\eeq
where both contributions can be obtained from (\ref{eq10}) by including 
$\tilde{V}^{J\pi(C)}_{c,c'}$ or $V^{J\pi(N)}_{c,c'}$ independently. 

Equation (\ref{eq10}) does not treat the Coulomb and nuclear interactions on an equal footing.
An alternative expression for the scattering matrix is based on solutions of the uncoupled system
\beq
&&\bigl(T_L+ V^{J\pi}_{c,c}(R)+E_{c}-E \bigr) \overline{g}^{J\pi}_{c}(R)=0,
\label{eq12b}
\eeq
where the associated (single-channel) scattering matrices are denoted as $\overline{U}^{J\pi}_{c}$. Then
the total scattering matrix can be written as
\beq
&&U^{J\pi}_{\omega L_{\omega},\alpha L}=\overline{U}^{J\pi}_{c} \delta_{\omega \alpha}\delta_{L_{\omega}L }+
 \nonumber \\
&&\frac{i}{2\hbar} \int_0^{\infty} \overline{g}^{J\pi}_{\alpha L}(R)
\sum_{\alpha' L'} \overline{V}^{J\pi}_{\alpha L,\alpha' L'}(R)
g^{J\pi}_{\omega L_{\omega},\alpha' L'}(R)dR,
\label{eq12c}
\eeq
where potential $\overline{V}^{J\pi}_{c,c'}(R)$ is now defined by
\beq
\overline{V}^{J\pi}_{c,c'}(R)=V^{J\pi}_{c,c'}(R)-V^{J\pi}_{c,c}(R) \delta_{c c'}.
\label{eq12d}
\eeq
In other words, potentials $\overline{V}^{J\pi}_{c,c'}(R)$ only contain coupling terms. In this method,
referred to as 'Method 2", the Coulomb and nuclear potentials are considered in a symmetric way. The 
scattering matrix is decomposed as
\beq
U^{J\pi}_{c,c'}=U^{J\pi(C2)}_{c,c'}+U^{J\pi(N2)}_{c,c'}.
\label{eq13b}
\eeq
Notice that the scattering matrix can also be written in a fully symmetric way, involving the
full CDCC wave functions $g^{J\pi}(R)$ only \cite{HM86}, but this is beyond the scope of the 
present work.

Definitions (\ref{eq10}) and (\ref{eq12c}) are exact, and stem from the Lippmann-Schwinger equation 
[notice that the scattering matrix is implicitly present in the r.h.s.\ through
the wave functions $g^{J\pi}(R)$]. On the one hand, they can be used to test the consistency of the calculation. The direct calculation of the scattering matrix and of its equivalent integral definitions (\ref{eq10}) and 
(\ref{eq12c}) are
indeed based on different approaches, and the equivalence represents a strong test of the calculation.
On the other hand, the integral definitions, although more complicated since they require the scattering
wave functions, permit a separation between the nuclear and Coulomb amplitudes. They are also used
to develop numerical techniques for scattering states \cite{RMP80b}.

It is interesting to observe that if the full radial wave function $g^{J\pi}(R)$ in Eq.\ (\ref{eq12c}) is approximated by the one-channel wave function  $\overline{g}^{J\pi}_{\omega{L_{\omega}}}(R) \delta_{\omega \alpha}\delta_{{L_{\omega}}L}$ of Eq. (\ref{eq12b}), we obtain the DWBA expression for non-diagonal part of the scattering matrix, 
\beq
&&U^{J\pi}_{\omega L_{\omega},\alpha L} - \overline{U}^{J\pi}_{\alpha L} \delta_{\omega \alpha}\delta_{L_{\omega}L }\approx 
U^{J\pi{\rm (DWBA)}}_{\omega L_{\omega},\alpha L}\nonumber\\
&&U^{J\pi{\rm (DWBA)}}_{\omega L_{\omega},\alpha L}=  \nonumber \\ 
&&\frac{i}{2\hbar}\int_0^{\infty} \overline{g}^{J\pi}_{\omega L_{\omega}}(R)\overline{V}^{J\pi}_{\alpha L,\omega L_{\omega}}(R)
\overline{g}^{J\pi}_{\alpha L}(R)dR.
\label{eq14b}
\eeq
The DWBA definition (\ref{eq14b}) is linear in the coupling potentials, as the one-channel radial wave functions do not contain 
coupling effects, except the average optical potential which generates the elastic scattering matrix $\overline{U}^{J\pi}_{c}$. 
The DWBA method is normally used for very weak coupling effects, such as in reactions where the $Q$-value is large, 
found in tightly bound nuclei. In reactions involving weakly bound nuclei, however, the DWBA approximation is not expected to be accurate,
since couplings to the continuum are important.

\subsection{Breakup cross sections}
For a zero-spin target, the total breakup cross section is defined from the scattering matrices as
\beq
\sigma_{\rm BU}(E)=\frac{\pi}{k_{\omega}^2}\sum_{J\pi} \sum_{\alpha L L_{\omega}}\frac{2J+1}{2I_{\omega}+1}
\bigl\vert U^{J\pi}_{\omega L_{\omega},\alpha L}(E) \bigr\vert ^2,
\label{eq7}
\eeq
where $I_{\omega}$ is the spin of the projectile, and where index $\alpha$ runs over continuum 
channels only.  This expression is a simple extension of inelastic cross sections, where the final 
channels $\alpha$ are bound states. 

To emphasize the role of the angular momentum, we recast the cross 
section (\ref{eq7}) as
\beq
\sigma_{\rm BU}(E)= \sum_{ L_{\omega}} \sigma_{\rm BU}(L_{\omega},E),
\label{eq8}
\eeq
with 
\beq
\label{eq9}
&&\sigma_{\rm BU}(L_{\omega},E)=\frac{\pi}{k_{\omega}^2} (2L_{\omega}+1) T_{L_{\omega}}(E),  \\
&& T_{L_{\omega}}(E)=\sum_{J\pi} \sum_{\alpha L}g_J
\bigl\vert U^{J\pi}_{\omega L_{\omega},\alpha L}(E) \bigr\vert ^2,  \\
&&g_J=\frac{2J+1}{(2I_{\omega}+1)(2L_{\omega}+1)}.
\eeq

As mentioned in the previous subsection, an element of the scattering matrix can be decomposed in
Coulomb and  nuclear components. Breakup cross sections obtained with these 
two terms are referred to as the Coulomb and nuclear cross sections, respectively. Definition (\ref{eq9}) 
is therefore written as
\beq
\sigma_{\rm BU}(L_{\omega},E)&=&\sigma_{\rm BU}^{(C)}(L_{\omega},E)
+ \sigma_{\rm BU}^{(N)}(L_{\omega},E) \nonumber \\
&&+ \sigma_{\rm BU}^{(int)}(L_{\omega},E),
\label{eq14}
\eeq
which is valid for decompositions (\ref{eq13}) and (\ref{eq13b}). Each partial-wave contribution 
involves the coefficients
\beq
&& T_{L_{\omega}}^{(C)}=\sum_{J\pi} \sum_{\alpha L}g_J \,
\bigl\vert U^{J\pi(C)}_{\omega L_{\omega},\alpha L} \bigr\vert ^2, \nonumber \\
&& T_{L_{\omega}}^{(N)}=\sum_{J\pi} \sum_{\alpha L}g_J \,
\bigl\vert U^{J\pi(N)}_{\omega L_{\omega},\alpha L}\bigr \vert ^2, \nonumber \\
&& T_{L_{\omega}}^{(int)}=2{\rm Re}\sum_{J\pi} \sum_{\alpha L}g_J \,
\bigl(U^{J\pi(N)}_{\omega L_{\omega},\alpha L}\bigr)^{\ast} \, U^{J\pi(C)}_{\omega L_{\omega},\alpha L}.
\label{eq15}
\eeq

In the CDCC method, the breakup cross section is therefore given by a multiple sum over several quantum numbers:
total spin and parity of the system, angular momentum and excitation level of the projectile. Our goal here
is not to reproduce experimental data, but to analyse in detail the various components of the breakup
cross section, and to discuss the importance of the nuclear and Coulomb contributions.

Finally, let us point out that the wordings 'nuclear' and 'Coulomb' components in the scattering matrix 
(\ref{eq13},\ref{eq13b}) are partly ambiguous. The reason is that, if the potentials can be clearly decomposed,
the scattering wave functions $g^{J\pi}(R)$ are obtained from the full Hamiltonian (\ref{eq4}), and are affected by the
nuclear interaction as well as by the Coulomb interaction. Also the existence of several possible ways
to decompose the breakup cross section raises the question of their equivalence. This will be analysed later for
the $\lipb$ breakup reaction.

In the literature, the separation between nuclear and Coulomb contributions is not well established. An approximate method, which does not make use of the wave functions has been proposed
 \cite{CGD15} to separate both terms. The idea is to solve the
coupled-channel system (\ref{eq4}) in two independent calculations.
Each calculation neglects, either nuclear couplings, or Coulomb couplings in (\ref{eq4}). They provide 
approximate Coulomb and nuclear breakup cross sections, respectively. This method avoids the more delicate calculation
of the scattering wave functions. However, it implicitly assumes that the coupling between the elastic and breakup channels
is weak enough to use a DWBA approximation. It what follows, this approximation will be referred to as the 'weak-coupling approximation'.

\subsection{Reaction cross sections}
Calculations of the reactions cross sections are often used to complement elastic-scattering
experiments, as they involve the same (elastic) elements of the scattering matrix. From a
theoretical point of view, the reaction cross section is directly related to the imaginary part
of the optical potential. If the potential is real, the reaction cross section exactly vanishes
\cite{CH13}.

In a multichannel theory, the reaction cross section is defined as
\beq
\sigma_{\rm R}(E)&=&\frac{\pi}{k_{\omega}^2}\sum_{J\pi} \sum_{ L L_{\omega}}\frac{2J+1}{2I_{\omega}+1} \nonumber \\
&&\times
\bigl(\delta_{ L L_{\omega}}- \vert U^{J\pi}_{\omega L_{\omega},\omega L}(E)\vert^2 \bigr).
\label{eqr1}
\eeq
In contrast with the breakup cross section, it only involves elements associated with the entrance channel
$\omega$.
This definition, as in Sec.\ \ref{sec3}.B, can be rewritten in an equivalent way as
\beq
\sigma_{\rm R}(E)= \sum_{ L_{\omega}}\sigma_{\rm R}( L_{\omega},E),
\label{eqr2}
\eeq
with
\beq
\sigma_{\rm R}( L_{\omega},E)&=&\frac{\pi}{k_{\omega}^2}(2L_{\omega}+1) \nonumber \\
&& \times \sum_{J\pi L}
g_J \bigl(\delta_{ L L_{\omega}}-\vert U^{J\pi}_{\omega L_{\omega},\omega L}(E)\vert^2 \bigr).
\label{eqr3}
\eeq

Let us now discuss the separation between the nuclear and Coulomb contributions. In the literature, this
issue has been essentially addressed for breakup, but not for reaction cross sections. Using
decomposition (\ref{eq13}), a partial reaction cross section (\ref{eqr2}) can be written as
\beq
\sigma_{\rm R}( L_{\omega},E)&=&\sigma_{\rm R}^{(C)}( L_{\omega},E)+
\sigma_{\rm R}^{(N)}( L_{\omega},E)\nonumber \\
&&+
\sigma_{\rm R}^{(int)}( L_{\omega},E),
\label{eqr4}
\eeq
with the Coulomb and nuclear contributions defined by
\beq
\sigma_{\rm R}^{(C)}( L_{\omega}&,&E)=\frac{\pi}{k_{\omega}^2}(2L_{\omega}+1)\nonumber \\
&&\times \sum_{J\pi }g_J
\bigl( 1-\sum_{L} \vert \delta_{ L L_{\omega}}+ U^{J\pi(C)}_{\omega L_{\omega},\omega L}\vert^2 \bigr)
\label{eqr5}
\eeq
and
\beq
\sigma_{\rm R}^{(N)}( L_{\omega}&,&E)=\frac{\pi}{k_{\omega}^2}(2L_{\omega}+1)\nonumber \\
&&\times \sum_{J\pi }g_J
\bigl( 1-\sum_{L} \vert \delta_{ L L_{\omega}}+ U^{J\pi(N)}_{\omega L_{\omega},\omega L}\vert^2 \bigr).
\label{eqr6}
\eeq
The interference term is given by
\beq
\sigma_{\rm R}^{(int)}( L_{\omega}&,&E)=-\frac{2\pi}{k_{\omega}^2}(2L_{\omega}+1)\nonumber \\
&&   \times {\rm Re}\sum_{J\pi }g_J
 \sum_{L}  \bigl( U^{J\pi(C)}_{\omega L_{\omega},\omega L}\bigr)^{\ast} U^{J\pi(N)}_{\omega L_{\omega},\omega L}.
\label{eqr7}
\eeq

Notice that a slightly different definition of the reaction cross section exists in the literature.
In Refs.\ \cite{Th88,TN09}, the reaction cross section is defined as
\beq
\sigma_{\rm R}(E)&=&\frac{\pi}{k_{\omega}^2}\sum_{J\pi} \sum_{L_{\omega}}\frac{2J+1}{2I_{\omega}+1} \nonumber \\
&&\times
\bigl(1- \vert U^{J\pi}_{\omega L_{\omega},\omega L_{\omega}}(E)\vert^2 \bigr).
\label{eqr8}
\eeq
Definitions (\ref{eqr1}) and (\ref{eqr8}) are equivalent for a projectile with $I_{\omega}=0$ or
$I_{\omega}=1/2$ (in that case the summation over $L_{\omega}$ is limited to a single value), but
differences may arise when $I_{\omega} \ge 1$. Using  (\ref{eqr8}) assumes that reorientation channels
(i.e.\ $L_{\omega} \ne L$ in (\ref{eqr1})) contribute to the reaction cross section, although they do enter
the definition of the elastic cross section. Another argument is in favor of (\ref{eqr1}). If the potentials are real, one expects the reaction cross section to be exactly zero. This is true with Eq.\ (\ref{eqr1}), owing to the
unitarity of the scattering matrix, but not with Eq.\ (\ref{eqr8}). The difference between both definitions
will be illustrated in Sec.\ \ref{sec5}.B.

\subsection{Pure Coulomb breakup}
In the previous subsections, the Coulomb breakup amplitude is partly affected by the nuclear interaction, since
the projectile+target wave function is obtained from the full Hamiltonian. Going further is the 'pure
Coulomb' approximation, where this scattering wave function is assumed to be a Coulomb wave. 

At large distances $R$, the three-body Coulomb interaction in (\ref{eq1b}) can be written in a multipole expansion as
\beq
V_C(\pmb{R},\pmb{r})\longrightarrow 4\pi Z_t e \sum_{\lambda}
\frac{\hat{\lambda}}
{R^{\lambda+1}}\bigl[{\cal M}^E_{\lambda}(\pmb{r})\otimes Y_{\lambda}(\Omega_R) \bigr]^0.
\label{eqc1}
\eeq
We define $\hat{x}=(2x+1)^{1/2}$, and we use $\pmb{\xi}=\pmb{r}$, the internal coordinate of the two-body projectile. 
The electric operator ${\cal M}^E_{\lambda}(\pmb{r})$ is associated with the projectile,
and is given by
\beq
{\cal M}^E_{\lambda \mu}(\pmb{r})
=e\biggl[ Z_1 \biggl( -\frac{A_2}{A}\biggr)^{\lambda} + Z_2 \biggl( \frac{A_1}{A}\biggr)^{\lambda} \biggr]
r^{\lambda}Y_{\lambda}^{\mu}(\Omega_r),
\label{eqc2}
\eeq
where the masses and charges of the fragments are denoted as $(A_1,A_2)$ and $(Z_1e,Z_2e)$, respectively.

In the framework of the Coulomb excitation \cite{AW75}, the wave functions are assumed to be Coulomb functions.
Within this assumption, a non-diagonal element of the scattering
matrix is written as
\beq
U^{J\pi(C)}_{\alpha L,\alpha' L'}&=&-2iZ_te\frac{\mu}{\hbar^2 (k_{\alpha}k_{\omega})^{1/2}} 
\sum_{\lambda} C^{\lambda}_{\alpha L,\alpha' L'}\nonumber \\
&& \times  
\langle \Phi^{j}_{0,n} \parallel {\cal M}^E_{\lambda} \parallel \Phi^{j'}_{0,n'} \rangle
I^{\lambda}_{\alpha L, \alpha' L'},
\label{eqc3}
\eeq
where the geometrical coefficients $C^{\lambda}_{\alpha L,\alpha' L'}$ are given by
\beq
C^{\lambda}_{\alpha L,\alpha' L'}&=&(-1)^{j+L'+J+\lambda}\hat{j}\hat{L}
\nonumber \\
&& \times \langle Y_L \parallel Y_{\lambda} \parallel Y_{L'} \rangle
\begin{Bmatrix}
	J&L&j \\
	\lambda & j'& L'
\end{Bmatrix}.
\label{eqc4}
\eeq
In Eq.\ (\ref{eqc3}), the Coulomb integrals are defined as
\beq
I^{\lambda}_{\alpha L, \alpha' L'}=4\pi
\int F_{L}(k_{\alpha}R)\, \frac{1}{R^{\lambda+1}}\, F_{ L'}(k_{\alpha'}R)\,  dR .
\label{eqc5}
\eeq

Integrals (\ref{eqc5}) are known to present several convergence difficulties: (i) large $R$-values are 
needed since the integrand presents fast oscillations and a slow decrease; (ii) the integrals
slowly converge with $L$ and, in general, large values are necessary to achieve a good accuracy on the
Coulomb breakup cross sections. Specific techniques have been developed to address these issues \cite{ABH56,OIK02}
(notice that different conventions exist for the normalization).

The Coulomb breakup cross section is then computed with the general definition (\ref{eq7}). The
Coulomb assumption, in practice, is not exactly satisfied. However, it can be used for a qualitative
description of the breakup process. Properties of the projectile and of the relative motion are then
factorized. The Coulomb integral does not vary much when the breakup energy $E_{\alpha'}$ changes. The main
dependence on this energy therefore comes for the electromagnetic matrix elements inside the projectile.
These matrix elements are associated with the reduced transition probabilities
\beq
B(E\lambda,j'n' \rightarrow jn)=\frac{2j+1}{2j'+1}
\bigl\vert \langle \Phi^{j}_{0,n} \parallel {\cal M}^E_{\lambda} \parallel \Phi^{j'}_{0,n'} \rangle \bigr\vert^2,
\label{eq_be}
\eeq

\section{Application to the $\lipb$ system}
\label{sec4}
\subsection{Conditions of the calculations}
For the $\li$ description, we adopt the
$\at$ potential of Ref.\ \cite{BM88}, defined as
\beq
V_0(r)=-(v_0+2\alpha v_{\ell s}\pmb{\ell}\cdot \pmb{s})
\exp(-\alpha r^2)+V_C(r),
\label{eq21}
\eeq
with $\alpha=0.15747$ fm$^{-2}$, $v_0=83.78$ MeV, and $v_{\ell s}=1.003$ MeV. The Coulomb potential
$V_C(r)$ is taken as the potential of an uniformly charged sphere with a radius $r_C=3.095$ fm. The $\at$ 
potential (\ref{eq21}) reproduces very well most properties of $\li$ bound states: energies, quadrupole
moment, $B(E2)$ value, radius, etc. The theoretical energy of the $7/2^-$ narrow resonance is 2.1 MeV, in
fair agreement with experiment (2.19 MeV). 

The Pauli principle between $\alpha$ and $t$ is simulated by
additional bound states corresponding to the Pauli forbidden states (fs), present in microscopic theories
\cite{Ho77}.
The potential contains one fs for $\ell=1,2$, and two fs for $\ell=0$. These states are not included in the
CDCC basis (\ref{eq3}).

The $\at$ wave functions (\ref{eq1}) are expanded over a Gauss-Laguerre basis \cite{Ba15} with 40 functions,
and a scaling factor $h=0.4$ fm. Up to $\emax=15$ MeV, the numbers of pseudostates are 13,12,13,13 for $\ell=0,1,2,3$, respectively.

\begin{figure}[htb]
	\begin{center}
		\epsfig{file=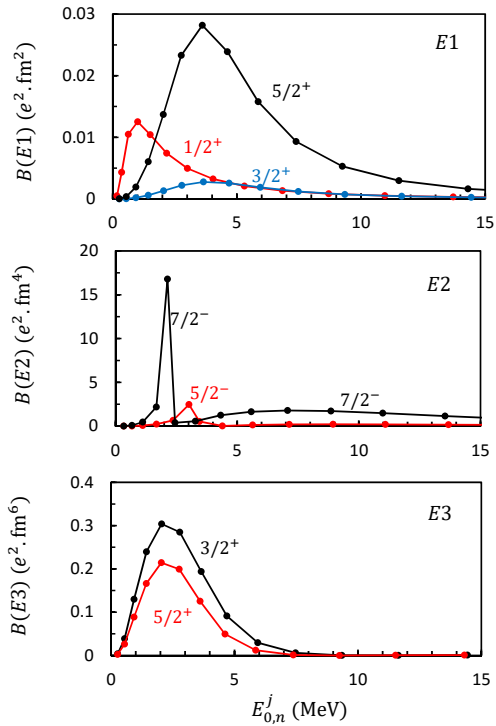,width=6.5cm}
		\caption{Electric transition probabilities (\ref{eq_be}) from the $3/2^-$ ground state as a function
			of the $\at$ continuum energy. Labels represent the final angular momentum $j$. For $E2$ transitions, the contributions of $j=1/2^-$ and of
			$j=3/2^-$ are negligible at the scale of the figure. The dots represent the (discrete) pseudostate energies
			$E^{j}_{0,n}$. The lines are to guide the eye.}
		\label{fig_be1}
	\end{center}
\end{figure}

In Fig.\ \ref{fig_be1}, we show the electric transition probabilities between the initial ground state
$j'=3/2^-$ and the various continuum states $j$. 
As mentioned before, these quantities are relevant for the breakup
process. In particular, the maxima are expected at similar energies. Transitions to $j=7/2^-$ show the
narrow resonance at 2.19 MeV, which is reasonably well approximated by a single eigenvalue. In this case,
$E^{7/2-}_{0,1}$ is a fair approximation of the resonance energy. Other partial waves present
broad peaks, which are described by a superposition of several pseudostates.

The model is complemented by $\alpha$+$^{208}$Pb and $t$+$^{208}$Pb optical potentials, which are taken from
Refs.\ \cite{AOR09} and \cite{PRS09}, respectively. Both potentials are expressed as Woods-Saxon form
factors, and involve volume and surface absorption terms. As mentioned previously, our goal here is not to fit
any data, but to analyse theoretical breakup and reaction cross sections. Therefore, no renormalization factors are
introduced in the optical potentials. The height of the Coulomb barrier is 27.7 MeV,
and its radius is 11.9 fm.

\begin{figure}[htb]
	\begin{center}
		\epsfig{file=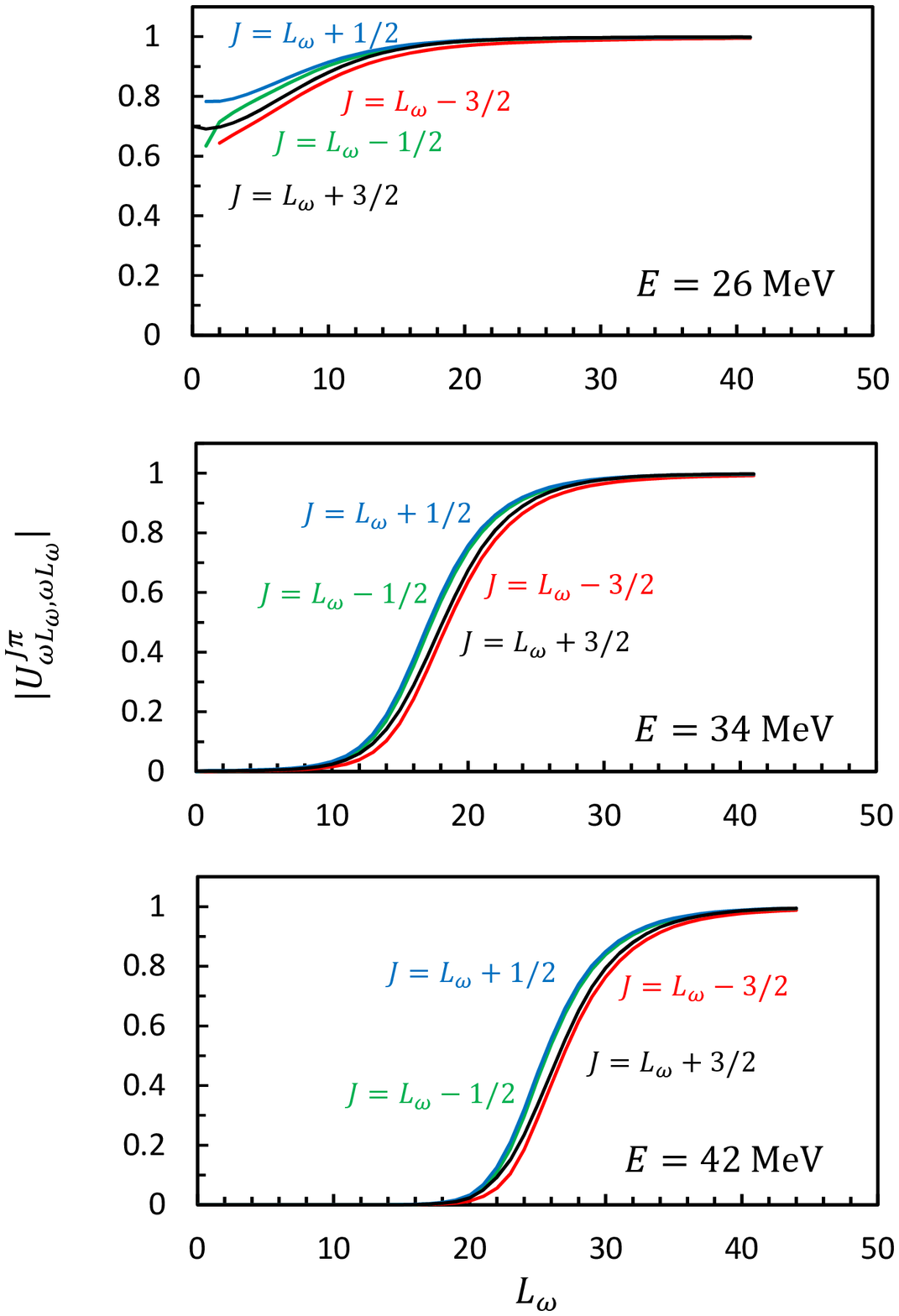,width=7cm,clip=}
		\caption{Diagonal amplitudes $\vert U^{J\pi}_{\omega L,\omega L}\vert$ at $E=26$, 34 and 42 MeV. The colors correspond to the four possibilities of $J$ values.}
		\label{fig_smat}
	\end{center}
\end{figure}

The scattering matrices are fundamental characteristics of the
scattering process. The diagonal elements (modulus) of the scattering matrix are shown in Fig.\ \ref{fig_smat} at three typical energies,
below and above the Coulomb barrier: 26, 34 and 42
MeV (throughout the text, the center-of-mass $\lipb$ scattering energy is denoted as $E$). According to the spin $j=3/2^-$ of the $\li$ ground state, 4 $J$-values are possible:
$J=L_{\omega}-3/2,L_{\omega}-1/2,L_{\omega}+1/2$ and $L_{\omega}+3/2$. As usual, the diagonal
scattering matrices are close to zero below a grazing angular momentum $L_g$, and tend to
unity at large $L_{\omega}$ values ($L_g\approx 26$ at $E=42$ MeV and $L_g\approx 17$ at $E=34$ MeV, the notion
of grazing angular momentum is not relevant below the Coulomb barrier). 

In the sharp-cutoff approximation 
($\vert U^{J\pi}_{\omega L,\omega L}\vert=0$ for $L\leq L_g$ and  
$\vert U^{J\pi}_{\omega L,\omega L}\vert=1$ for $L > L_g$) analytical approximation of
the elastic cross section can be derived \cite{Fr66}. The scattering matrices are also
helpful to interpret qualitatively inelastic or breakup cross sections \cite{Fr85} (see Sec.\ \ref{sec5}).

\subsection{Total breakup cross sections}
With the adopted conditions of calculations ($\emax=15$ MeV, $\lmax=3$), the model involves 89 pseudostates and two physical bound states. Considering that one channel contains in general several $L$ values, the size
of the coupled-channel system (\ref{eq4}) can be of the order of 300. This is efficiently solved with
the $R$-matrix method associated with propagation techniques \cite{DB10}.

\begin{figure}[htb]
	\begin{center}
		\epsfig{file=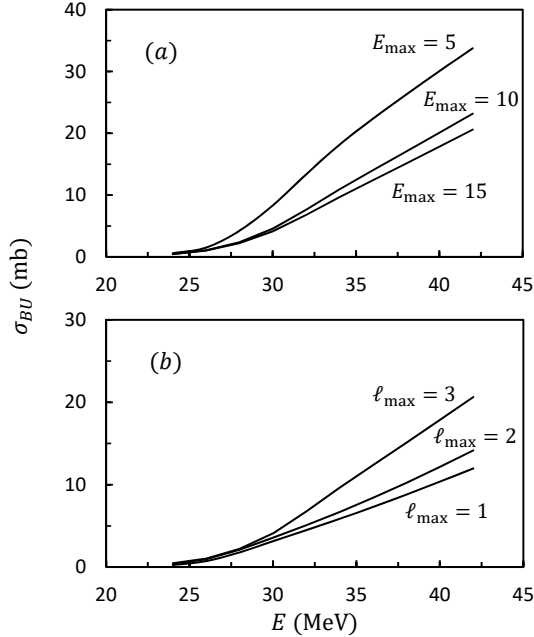,width=7cm}
		\caption{Convergence of the breakup cross section with $\emax$ (a) ($\emax=15$ MeV and $\emax=20$ MeV are
			indistinguishable)  and with $\lmax$ (b). In (a), $\lmax=3$ is used, and $\emax=15$ MeV is used in (b).}
		\label{fig_buconv}
	\end{center}
\end{figure}

We first present, in Fig.\ \ref{fig_buconv}, the total breakup cross sections, and analyze the convergence against the truncation parameters $\emax$ [Fig.\ \ref{fig_buconv}(a)] and $\lmax$ [Fig.\ \ref{fig_buconv}(b)]. These parameters
determine the number of pseudostates (index $\alpha$ in Eq.\ (\ref{eq7})), and therefore the $\at$ continuum. In
Fig.\ \ref{fig_buconv}(a), we fix $\lmax=3$, and vary the energy $\emax$. At least $\emax=10$ MeV is necessary
to achieve a fair convergence. In what follows, we use $\emax=15$ MeV. 

The sensitivity to $\lmax$ is presented in Fig.\ \ref{fig_buconv}(b). We observe a weak sensitivity below 30
MeV, which is slightly above the Coulomb barrier (27.7 MeV). However the role $\ell=3$ is obvious at higher
energies, and arises from the $7/2^-$ narrow resonance. A similar effect has been observed for the $^6$Li
breakup \cite{CDG16}. Throughout the text, we use $\lmax=3$, unless specified otherwise.

\subsection{Role of the $\at$ continuum}
In this subsection, our goal is to investigate the different contributions of the $\at$ continuum. We reformulate the breakup cross section (\ref{eq7}) as
\beq
\sigma_{\rm BU}(E)=\sum_{j,n} \sigma_{\rm BU}^j(E,E^j_{0,n}),
\label{eq22}
\eeq
where $j$ is associated with the $\at$ angular momentum and parity, and $E^j_{0,n}$ are the pseudostate energies.
The partial breakup cross section reads
\beq
\sigma^j_{\rm BU}(E,E^j_{0,n})=\frac{\pi}{k_{\omega}^2}\sum_{J\pi} \sum_{L, L_{\omega}}\frac{2J+1}{2I_{\omega}+1}
\bigl\vert U^{J\pi}_{\omega L_{\omega},jn L} \bigr\vert ^2.
\label{eq22b}
\eeq

\begin{figure}[htb]
	\begin{center}
		\epsfig{file=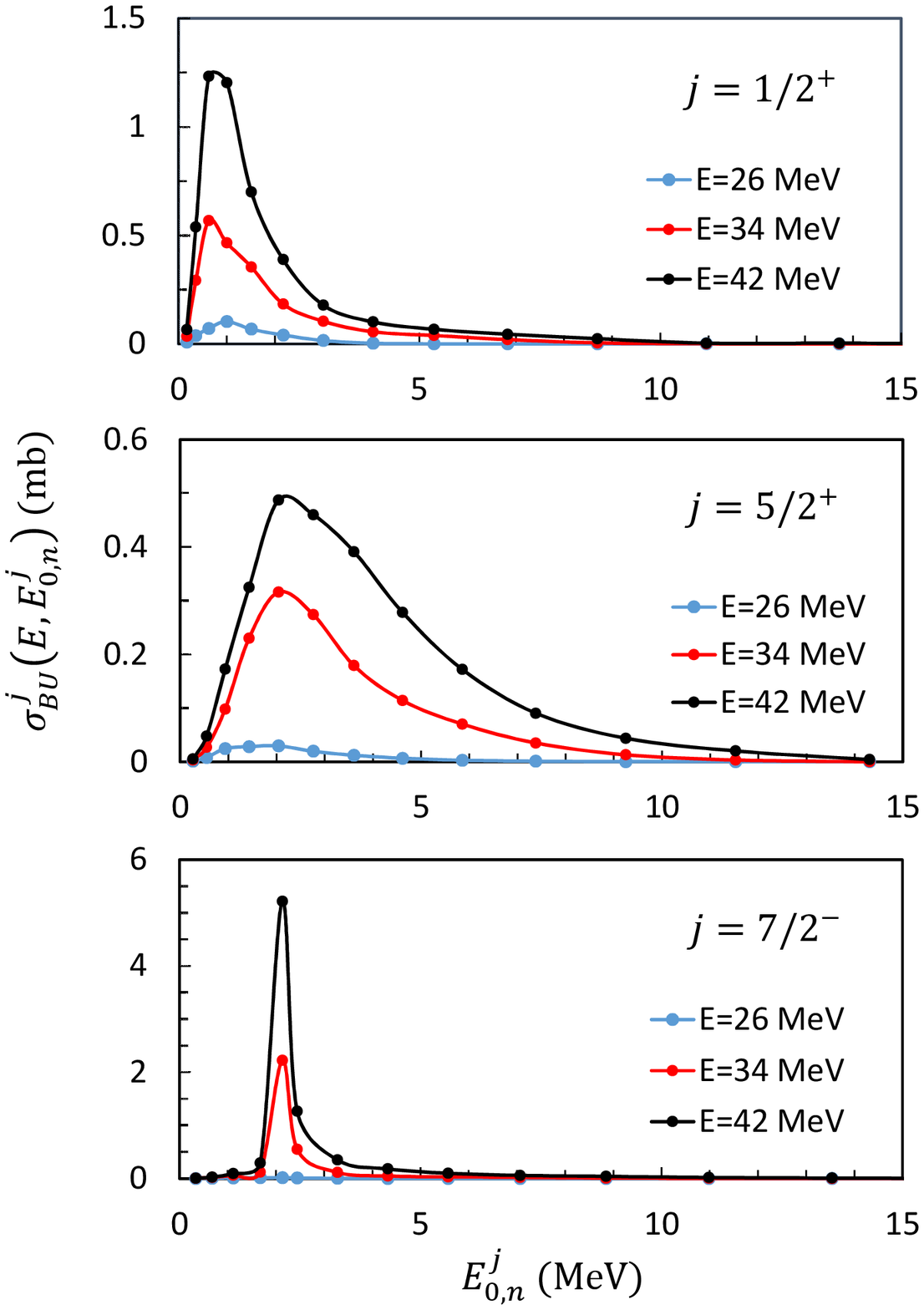,width=7cm}
		\caption{Breakup cross sections (\ref{eq22b}) for different partial waves $j$ of the $\at$ continuum. The $\lipb$
			energies are 26, 34 and 42 MeV. The dots represent the pseudostate energies, and the lines are to guide the eye.}
		\label{fig_etats}
	\end{center}
\end{figure}

In Fig.\ \ref{fig_etats}, we present the contributions of the most important partial waves $j=1/2^+, 3/2^-$,
and $7/2^-$. 
The cross section is analysed at three typical energies ($E=26,34,42$ MeV). Each cross section follows the shape 
of the $B(E\lambda)$ presented in  Fig.\ \ref{fig_be1}, with a maximum at low $\at$ energies. The $7/2^-$
resonance is clearly visible. The amplitudes vary with energy, as it is expected from barrier penetration effects.

\begin{figure}[htb]
	\begin{center}
		\epsfig{file=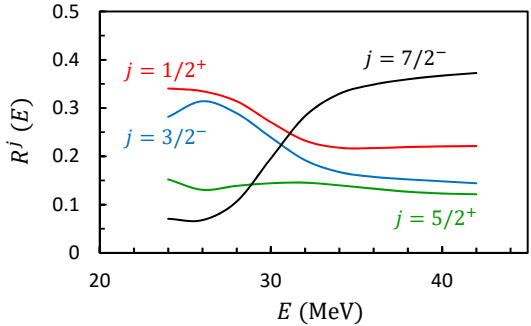,width=7cm}
		\caption{Contributions of the dominant $\at$ partial waves $j$
			to the total breakup cross section [see Eq.\ (\ref{eq_ratio})].}
		\label{fig_etats2}
	\end{center}
\end{figure}

Figure \ref{fig_etats2} provides the ratios of the different $\at$ partial waves, defined as
\beq
R^j(E)=\frac{\sum_n \sigma^j_{\rm BU}(E,E^j_{0n})}{\sigma_{\rm BU}(E)},
\label{eq_ratio}
\eeq
This definition involves a summation over the pseudostates in partial wave $j$. At low scattering energies, the dominant contribution stems from the lowest angular momenta $j=3/2^-\, (\ell=0)$ and $j=1/2^+\, (\ell=1)$, with a minor
contribution of $j=7/2^-\, (\ell=3)$. When the energy increases, however, the $7/2^-$ resonance is more and more
important. It starts being dominant around 31 MeV, which roughly corresponds to the barrier energy plus the
excitation energy of the $7/2^-$ resonance.

\subsection{Partial-wave analysis}
\begin{figure}[htb]
	\begin{center}
		\epsfig{file=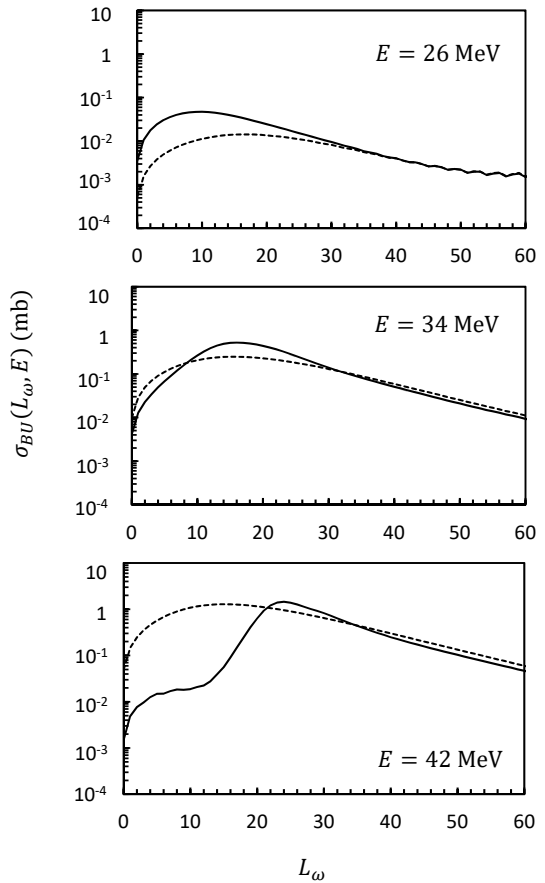,width=7cm}
		\caption{Angular-momentum distribution of the breakup cross sections at $E=26,34,42$ MeV. The solid
			lines represent the full CDCC calculation, while the dashed line are obtained at the 
			Coulomb approximation (\ref{eqc3}).}
		\label{fig_lom}
	\end{center}
\end{figure}

In this subsection, we analyze the angular-momentum distribution of the breakup cross sections.
The individual components (\ref{eq8}) are plotted in Fig.\ \ref{fig_lom} for three typical energies (solid lines).
We also present the pure Coulomb cross sections (dashed lines), obtained with the approximate matrix
elements (\ref{eqc3}). This provides a qualitative information on the Coulomb contribution, and does not
depend on any model assumption. At this stage, we do not consider the separation between nuclear and
Coulomb breakup. This will be done in the next section.

At large angular momenta, both distributions coincide since the centrifugal barrier strongly reduces the
role of the nuclear interaction, and only the Coulomb term remains. Notice that the exact matching starts
around $L_{\omega}\approx 80$ for $E=34$ MeV and around $L_{\omega}\approx 100$ for $E=42$ MeV.

Above the Coulomb barrier, i.e.\ when the grazing angular momentum $L_g$ satisfies the condition $L_g\gg 1$,
the strong-absorption approximation \cite{Fr85} can be used. In particular the 'Sopkovich prescription'
\cite{So62} provides a link between the CDCC scattering matrix $U^{J\pi}_{c,c'}$ and the Coulomb 
approximation $U^{J\pi(C)}_{c,c'}$ (\ref{eqc3}).
This prescription reads
\beq
U^{J\pi}_{c,c'}\approx U^{J\pi(C)}_{c,c'}\times  \bigl( \overline{U}^{J\pi}_c \times \overline{U}^{J\pi}_{c'}
\bigr)^{1/2},
\label{eq23}
\eeq
where $\overline{U}^{J\pi}_c$ is the single-channel scattering matrix. The consequence is that a breakup
element is expected to follow the $L$-dependence of the single-channel $S$-matrices. From Fig.\ \ref{fig_smat},
we may expect that, if the energy is high enough ($E=42$ MeV in our example), the breakup element is
strongly damped below the grazing angular momentum ($L_{g}\approx 26$ from Fig.\ \ref{fig_smat}). Consequently, the
dominant contribution in breakup cross sections comes from angular momenta around the grazing value, as for 
elastic scattering. Of course these properties are qualitative only, but provide a simple and natural
explanation to the more complicated CDCC breakup calculations.

\section{Nuclear and Coulomb contributions in $\lipb$}
\label{sec5}
\subsection{Breakup cross sections}
\begin{figure}[htb]
	\begin{center}
		\epsfig{file=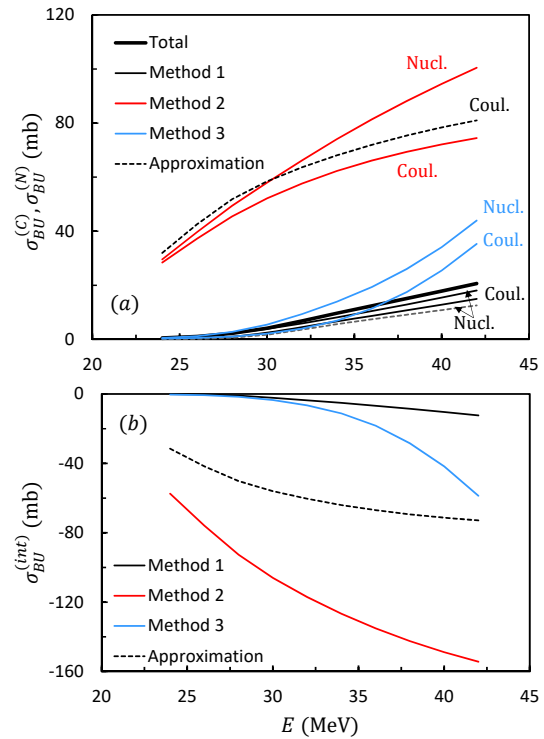,width=7cm}
		\caption{Nuclear and Coulomb (a) and interference (b) components of the breakup cross section.}
		\label{fig_bucn}
	\end{center}
\end{figure}

Our main goal in this work is to analyse the nuclear and Coulomb cross sections, as discussed in Sec.\ \ref{sec3}. 
The nuclear, Coulomb and interference contributions, as well as the total cross sections are presented in Fig.\ \ref{fig_bucn}.
This figure shows that methods 1 and 2 (Eqs.\ (\ref{eq13}) and (\ref{eq13b}), respectively) provide
rather different nuclear and Coulomb cross sections. Note, however, that the total breakup cross section (see also Fig.\ \ref{fig_buconv}) is,
up to the numerical precision, exactly reproduced. 
This large difference between both methods suggests that the separation of nuclear and Coulomb breakup is
ambiguous. The origin of this ambiguity comes from the fact that the wave functions depend
on the nuclear and Coulomb interactions simultaneously. The separation of both potentials, which is
unambiguous, is not sufficient to clearly separate the nuclear and Coulomb components of the breakup
cross section. With both methods, the nuclear and Coulomb contributions are close to each other, and interference terms are
important. Strong nuclear effects has been also pointed out in Ref.\ \cite{DLV98}.

In Fig.\ \ref{fig_bucn}, we also present the cross sections obtained with the weak-coupling approximation \cite{CGD15}
(dashed lines). This approximation is very close to method 1 for the nuclear breakup, and to method 2 for
the Coulomb breakup. The agreement is striking and difficult to explain quantitatively. Qualitatively, however,
the physical interpretation is as follows. In the weak-coupling approximation, one generates total wave functions
which differ from the exact wave functions since either the nuclear or the Coulomb potential is neglected.
In the weak-coupling approximation, the nuclear component is obtained by switching off all Coulomb couplings. This
means that the approximate wave function is expected to be more accurate at short distances. Therefore it is
more consistent with method 1, which contains diagonal nuclear potentials. An equivalent conclusion holds for
the approximate Coulomb cross section, which is expected to be more accurate with method 2. This is of course a
qualitative interpretation. The conclusion about the ambiguity, however, remains and, from our example, the notion 
of 'nuclear' or 'Coulomb' breakup is quite unclear.

\begin{figure}[htb]
	\begin{center}
		\epsfig{file=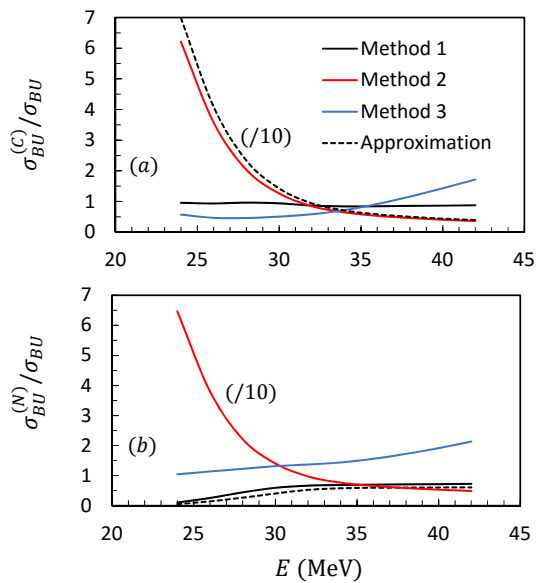,width=7cm}
		\caption{Fraction of Coulomb (a) and nuclear (b) breakup cross sections. The solid lines are obtained as described
			in the text, and the dashed lines represent the weak-coupling approximation.}
		\label{fig_ratio}
	\end{center}
\end{figure}

In order to visualize the low-energy cross sections, we display in Fig.\
\ref{fig_ratio} the ratio of the Coulomb [Fig.\ \ref{fig_ratio}(a)] and nuclear [Fig.\ \ref{fig_ratio}(b)] 
cross sections, divided by the total breakup cross section. Notice that the sum of these ratios is different from unity since 
interference terms (which can be negative) exist in definition (\ref{eq14}). Interferences increase with energy, as suggested by Fig.\ \ref{fig_bucn}.
Whereas method 1 gives ratios of the order of unity, method 2,
more symmetric in the treatment of the nuclear and Coulomb interactions, is strongly energy dependent, and
the ratios are much larger than 1 (for the scale of clarity the results have been scaled by 1/10 in the Figure).

Methods 1 and 2 are based on integral definitions of the scattering matrix. Although the nuclear and Coulomb components
can be clearly separated in the potential, their contributions to the wave functions cannot be disentangled.
This problem leads to the ambiguity discussed above. There is, however, an approach which is model
independent, and along the ideas used for elastic or inelastic scattering \cite{DD12b}, or even for
bremsstrahlung \cite{BSD91}.
Let us define a nuclear contribution by
\beq
U^{J\pi}_{c,c'}=U^{J\pi(C)}_{c,c'}+U^{J\pi(N3)}_{c,c'},
\label{eq24}
\eeq
where $U^{J\pi(C)}_{c,c'}$ is the so-called "pure-Coulomb" scattering matrix (\ref{eqc3}). In this way
the Coulomb term does not depend on the reaction model [it still depends on the projectile properties
through the $E\lambda$ matrix elements (\ref{eq_be})]. This method, hereafter referred to as "Method 3", is
illustrated in Figs.\ \ref{fig_bucn} and \ref{fig_ratio}. Its main advantage is to address the slow convergence of the CDCC
breakup cross sections. As shown in the $L$-distribution of Fig.\ \ref{fig_lom}, the convergence with angular
momentum is slow, in particular at low incident energies, but the matrix elements at large $L$ can be computed
separately by using specific techniques \cite{OIK02}. In this way, the full CDCC calculation is not necessary beyond a certain $L$ value, and can be accurately replaced by the Coulomb approximation. This technique was used in Ref.\ \cite{DD12b} for inelastic
scattering, and can be adapted to breakup reactions.

\begin{figure}[htb]
	\begin{center}
		\epsfig{file=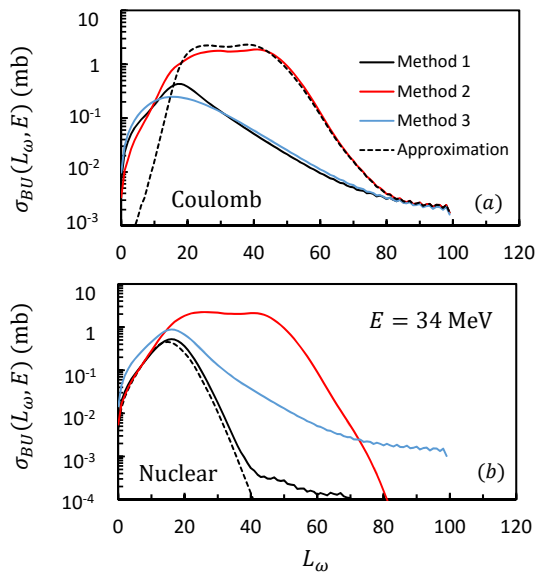,width=7cm}
		\caption{Partial-wave decomposition of the Coulomb (a) and nuclear (b) breakup cross section at $E=34$ MeV.}
		\label{fig_buj}
	\end{center}
\end{figure}

Finally we explore the $L$ dependence of the various methods. We select an intermediate energy $E=34$ MeV, and plot in 
Fig.\ \ref{fig_buj} the Coulomb (a) and nuclear (b) cross sections as a function of the angular momentum $L_{\omega}$.
As expected from the analysis of the total cross sections (Fig.\ \ref{fig_bucn}), the three methods
and their approximation provide different results. At low angular momenta, and at high angular momenta, the results are consistent with each other. However there is a wide window ($20 \lesssim L_{\omega} \lesssim 80$) where all
separation techniques provide different results. A similar conclusion is drawn at other energies, which 
confirms the ambiguity when trying to disentangle the Coulomb an nuclear components
in breakup cross sections.

\subsection{Reaction cross sections}
\begin{figure}[htb]
	\begin{center}
		\epsfig{file=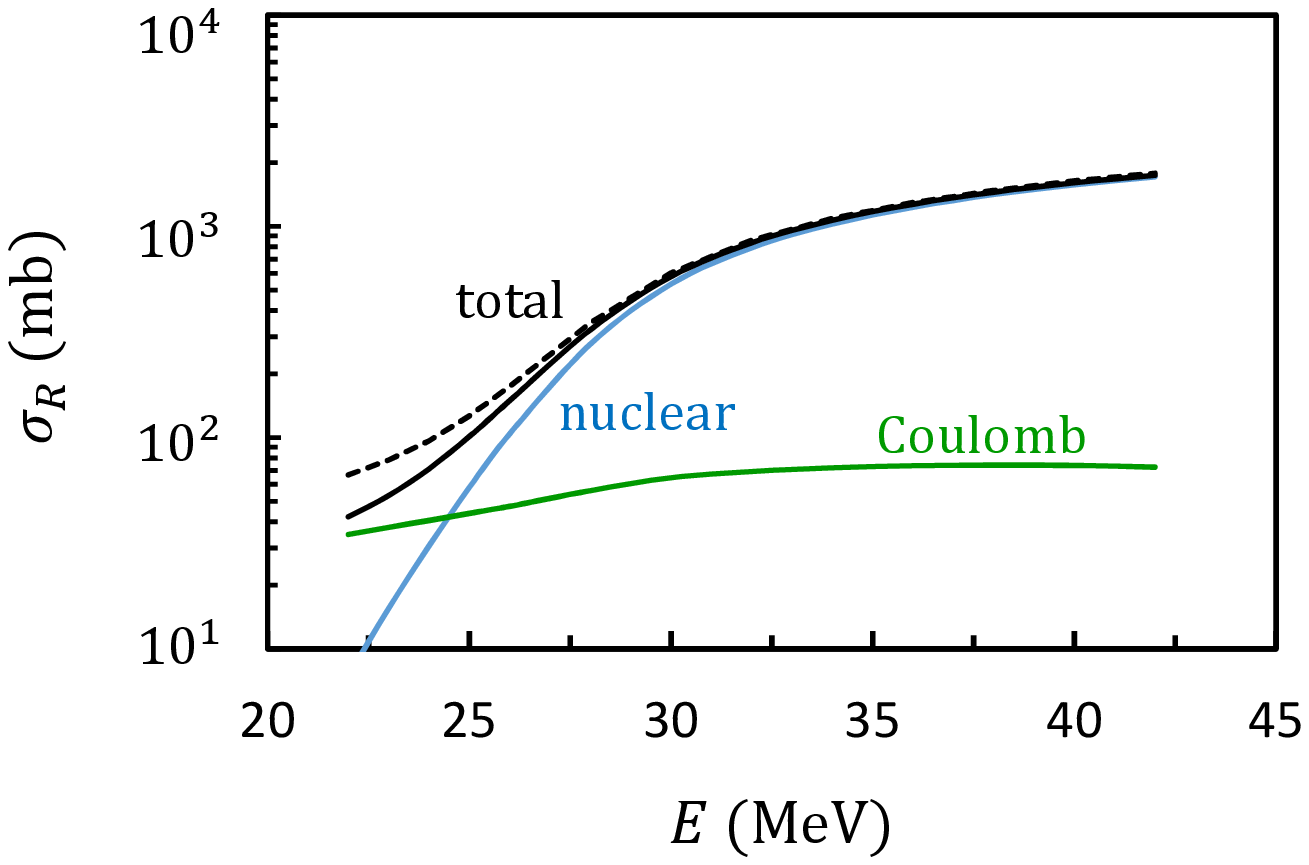,width=7cm}
		\caption{Reaction cross section (\ref{eqr2}) with the Coulomb (\ref{eqr5}) and nuclear (\ref{eqr6})
			contributions. The dashed line corresponds to the definition (\ref{eqr8}) of the reaction cross section.}
		\label{fig_reac}
	\end{center}
\end{figure}

This subsection is devoted to reaction cross sections. Figure \ref{fig_reac} presents the CDCC
results with the nuclear and Coulomb contributions. 
Here only method 1 can be used as method 2 involves a single-channel elements which cannot be directly
separated in nuclear and Coulomb terms.
In contrast with breakup, the reaction
cross section is dominated by nuclear effects above the Coulomb barrier. Around 25 MeV (i.e.\ 2 MeV
below the Coulomb barrier), however, both contributions are similar. Coulomb effects are dominant
below 25 MeV, an energy region where the reaction cross section is rather small. We present as a dashed
line the reaction cross section deduced from definition (\ref{eqr8}). In the present case, the difference
is negligible, except below the Coulomb barrier.

\begin{figure}[htb]
	\begin{center}
		\epsfig{file=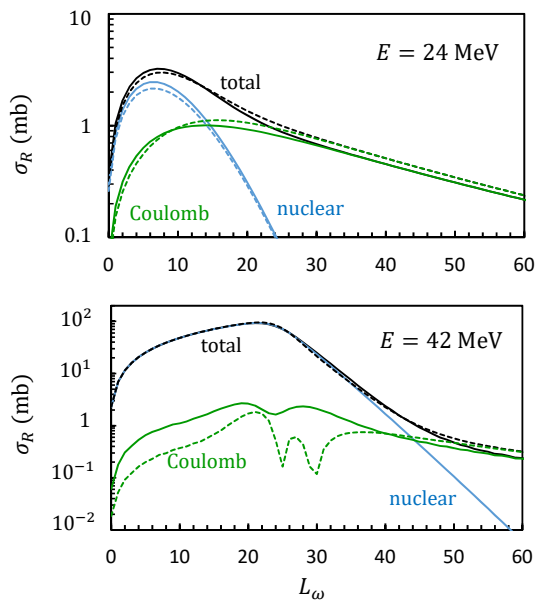,width=7cm}
		\caption{Partial-wave decomposition of the reaction cross section at $E=24$ MeV (top) and $E=42$ MeV
			(bottom). The Coulomb and nuclear contributions are shown in green and blue, respectively. The solid
			lines correspond to the full CDCC calculation, and the dashed lines to the single-channel
			approximation (at $E=42$ MeV, the nuclear terms are indistinguishable). The single-channel Coulomb
			term is negative for $25 \leq L_{\omega} \leq 29$, and the absolute value is shown.}
		\label{fig_reac_sep}
	\end{center}
\end{figure}

We extend our analysis to a partial-wave decomposition of the reaction cross section. The distribution
is displayed in Fig.\ \ref{fig_reac_sep}, with the individual Coulomb and nuclear components. As suggest
by Fig.\ \ref{fig_reac}, Coulomb effects are negligible above the Coulomb barrier (here we take $E=42$ MeV). Even
if the $L$-convergence is slow, the contribution of low partial waves (up to $L_{\omega}\approx 40$) is
strongly dominant. This is less true at 24 MeV, where the nuclear term is present in a narrow range 
(up to $L_{\omega}\approx 15$). In these circumstances, the convergence with angular momentum must be considered very carefully.

For the sake of completeness, we show in Fig.\ \ref{fig_reac_sep} the various cross sections in the
single-channel approximation (dashed lines). Clearly the role of the $\at$ continuum is minor.

\section{Conclusion}
\label{sec6}
The main goal of this work was to address, in the CDCC framework, the separation between Coulomb and nuclear components
in breakup cross sections. The CDCC method provides an exact solution of the three-body problem, and is therefore well adapted to
breakup reactions.
In contrast with the total cross section, which only requires the
scattering matrices, the individual contributions are based on integrals using the scattering wave
functions and the potentials. 
We find that, even at low energies, nuclear effects are not negligible, as suggested long ago by Dasso {\sl et al}.\ \cite{DLV98}.
These authors showed that assuming a Coulomb dissociation, as it is usually done, is questionable. Our work confirms this
conclusion, but goes even further. At high angular momenta, the situation is fairly clear: the nuclear term is negligible,
and the  Coulomb contribution is well defined. This is, however, not true at low angular
momenta. Not only nuclear effects are important, but the separation itself depends on the method used.
This is not very surprising since $(i)$  both
contributions contain the wave function, which depends on the full interaction; $(ii)$ there are several
methods to separate both components, which provide significantly different results.
These results therefore suggest that the wordings 'Coulomb' and 'nuclear' are partly ambiguous, and that extracting information from
dissociation data cannot be done by neglecting {\sl a priori} nuclear effects.

Our test of the weak-coupling approximation provides rather striking results. For $\lipb$, it resembles nuclear 
breakup with method 1, and Coulomb breakup with method 2. A qualitative interpretation has been suggested,
but it is difficult to go beyond this simple discussion. Our comparison confirms that Coulomb and nuclear cross sections obtained in this way are questionable.

A similar decomposition has been performed on reaction cross sections. Above the Coulomb barrier,
the cross section is essentially nuclear, and Coulomb effects are negligible. We have discussed
the definition itself of the reaction cross section, which may differ in the literature for 
projectiles with a non-zero spin. In the present case, this effect is limited above the Coulomb barrier, 
and probably within the reliability of the model assumptions.

The conclusion drawn from the $\lipb$ system should be rather general. However, specific effects may be
present in other reactions. In particular it would be interesting to perform similar analyses for projectiles
with $N/Z=1$, where Coulomb effects are expected to be weaker. The extension of the present framework to
three-body projectiles also deserves further work.

\section*{Acknowledgments}
One of us (P. D.) is grateful to M. Rodr\'iguez-Gallardo for her assistance in the numerical calculations, and
to I. J. Thompson for discussions about the reaction cross section.
This text presents research results of the IAP programme P7/12 initiated by the Belgian-state 
Federal Services for Scientific, Technical and Cultural Affairs. 
P. D. is Directeur de Recherches of F.R.S.-FNRS, Belgium. Partial support from the Brazilian funding agencies, CNPq, 
FAPESP, and FAPERJ is also acknowledged. M. S. H. acknowledges support from the CAPES/ITA Senior Visiting Professor 
Fellowship Program.


\begin{thebibliography}{49}%
	\makeatletter
	\providecommand \@ifxundefined [1]{%
		\@ifx{#1\undefined}
	}%
	\providecommand \@ifnum [1]{%
		\ifnum #1\expandafter \@firstoftwo
		\else \expandafter \@secondoftwo
		\fi
	}%
	\providecommand \@ifx [1]{%
		\ifx #1\expandafter \@firstoftwo
		\else \expandafter \@secondoftwo
		\fi
	}%
	\providecommand \natexlab [1]{#1}%
	\providecommand \enquote  [1]{``#1''}%
	\providecommand \bibnamefont  [1]{#1}%
	\providecommand \bibfnamefont [1]{#1}%
	\providecommand \citenamefont [1]{#1}%
	\providecommand \href@noop [0]{\@secondoftwo}%
	\providecommand \href [0]{\begingroup \@sanitize@url \@href}%
	\providecommand \@href[1]{\@@startlink{#1}\@@href}%
	\providecommand \@@href[1]{\endgroup#1\@@endlink}%
	\providecommand \@sanitize@url [0]{\catcode `\\12\catcode `\$12\catcode
		`\&12\catcode `\#12\catcode `\^12\catcode `\_12\catcode `\%12\relax}%
	\providecommand \@@startlink[1]{}%
	\providecommand \@@endlink[0]{}%
	\providecommand \url  [0]{\begingroup\@sanitize@url \@url }%
	\providecommand \@url [1]{\endgroup\@href {#1}{\urlprefix }}%
	\providecommand \urlprefix  [0]{URL }%
	\providecommand \Eprint [0]{\href }%
	\providecommand \doibase [0]{http://dx.doi.org/}%
	\providecommand \selectlanguage [0]{\@gobble}%
	\providecommand \bibinfo  [0]{\@secondoftwo}%
	\providecommand \bibfield  [0]{\@secondoftwo}%
	\providecommand \translation [1]{[#1]}%
	\providecommand \BibitemOpen [0]{}%
	\providecommand \bibitemStop [0]{}%
	\providecommand \bibitemNoStop [0]{.\EOS\space}%
	\providecommand \EOS [0]{\spacefactor3000\relax}%
	\providecommand \BibitemShut  [1]{\csname bibitem#1\endcsname}%
	\let\auto@bib@innerbib\@empty
	\bibitem [{\citenamefont {Bertulani}\ \emph {et~al.}(2001)\citenamefont
		{Bertulani}, \citenamefont {Hussein},\ and\ \citenamefont
		{M\"unzenberg}}]{BHM01}%
	\BibitemOpen
	\bibfield  {author} {\bibinfo {author} {\bibfnamefont {C.~A.}\ \bibnamefont
			{Bertulani}}, \bibinfo {author} {\bibfnamefont {M.~S.}\ \bibnamefont
			{Hussein}}, \ and\ \bibinfo {author} {\bibfnamefont {G.}~\bibnamefont
			{M\"unzenberg}},\ }\href@noop {} {\emph {\bibinfo {title} {Physics of
				Radioactive Beams}}}\ (\bibinfo  {publisher} {Nova Science Publishers, New
		York},\ \bibinfo {year} {2001})\BibitemShut {NoStop}%
	\bibitem [{\citenamefont {Tanihata}\ \emph {et~al.}(2013)\citenamefont
		{Tanihata}, \citenamefont {Savajols},\ and\ \citenamefont {Kanungo}}]{TSK13}%
	\BibitemOpen
	\bibfield  {author} {\bibinfo {author} {\bibfnamefont {I.}~\bibnamefont
			{Tanihata}}, \bibinfo {author} {\bibfnamefont {H.}~\bibnamefont {Savajols}},
		\ and\ \bibinfo {author} {\bibfnamefont {R.}~\bibnamefont {Kanungo}},\
	}\href@noop {} {\bibfield  {journal} {\bibinfo  {journal} {Prog. Part. Nucl.
			Phys.}\ }\textbf {\bibinfo {volume} {68}},\ \bibinfo {pages} {215} (\bibinfo
	{year} {2013})}\BibitemShut {NoStop}%
\bibitem [{\citenamefont {Blumenfeld}\ \emph {et~al.}(2013)\citenamefont
	{Blumenfeld}, \citenamefont {Nilsson},\ and\ \citenamefont
	{Van~Duppen}}]{BNV13}%
\BibitemOpen
\bibfield  {author} {\bibinfo {author} {\bibfnamefont {Y.}~\bibnamefont
		{Blumenfeld}}, \bibinfo {author} {\bibfnamefont {T.}~\bibnamefont {Nilsson}},
	\ and\ \bibinfo {author} {\bibfnamefont {P.}~\bibnamefont {Van~Duppen}},\
}\href@noop {} {\bibfield  {journal} {\bibinfo  {journal} {Physica Scripta}\
}\textbf {\bibinfo {volume} {2013}},\ \bibinfo {pages} {014023} (\bibinfo
{year} {2013})}\BibitemShut {NoStop}%
\bibitem [{\citenamefont {L\'epine-Szily}\ \emph {et~al.}(2014)\citenamefont
	{L\'epine-Szily}, \citenamefont {Lichtenth\"aler},\ and\ \citenamefont
	{Guimar\~aes}}]{LLG14}%
\BibitemOpen
\bibfield  {author} {\bibinfo {author} {\bibfnamefont {A.}~\bibnamefont
		{L\'epine-Szily}}, \bibinfo {author} {\bibfnamefont {R.}~\bibnamefont
		{Lichtenth\"aler}}, \ and\ \bibinfo {author} {\bibfnamefont {V.}~\bibnamefont
		{Guimar\~aes}},\ }\href@noop {} {\bibfield  {journal} {\bibinfo  {journal}
		{Eur. Phys. J. A}\ }\textbf {\bibinfo {volume} {50}},\ \bibinfo {pages} {128}
	(\bibinfo {year} {2014})}\BibitemShut {NoStop}%
\bibitem [{\citenamefont {Tanihata}(2016)}]{Ta16}%
\BibitemOpen
\bibfield  {author} {\bibinfo {author} {\bibfnamefont {I.}~\bibnamefont
		{Tanihata}},\ }\href {\doibase 10.1140/epjp/i2016-16090-x} {\bibfield
	{journal} {\bibinfo  {journal} {Eur. Phys. J. Plus}\ }\textbf {\bibinfo
		{volume} {131}},\ \bibinfo {pages} {1} (\bibinfo {year} {2016})}\BibitemShut
{NoStop}%
\bibitem [{\citenamefont {Kamimura}\ \emph {et~al.}(1986)\citenamefont
	{Kamimura}, \citenamefont {Yahiro}, \citenamefont {Iseri}, \citenamefont
	{Sakuragi}, \citenamefont {Kameyama},\ and\ \citenamefont {Kawai}}]{KYI86}%
\BibitemOpen
\bibfield  {author} {\bibinfo {author} {\bibfnamefont {M.}~\bibnamefont
		{Kamimura}}, \bibinfo {author} {\bibfnamefont {M.}~\bibnamefont {Yahiro}},
	\bibinfo {author} {\bibfnamefont {Y.}~\bibnamefont {Iseri}}, \bibinfo
	{author} {\bibfnamefont {S.}~\bibnamefont {Sakuragi}}, \bibinfo {author}
	{\bibfnamefont {H.}~\bibnamefont {Kameyama}}, \ and\ \bibinfo {author}
	{\bibfnamefont {M.}~\bibnamefont {Kawai}},\ }\href@noop {} {\bibfield
	{journal} {\bibinfo  {journal} {Prog. Theor. Phys. Suppl.}\ }\textbf
	{\bibinfo {volume} {89}},\ \bibinfo {pages} {1} (\bibinfo {year}
	{1986})}\BibitemShut {NoStop}%
\bibitem [{\citenamefont {Rawitscher}(1974)}]{Ra74b}%
\BibitemOpen
\bibfield  {author} {\bibinfo {author} {\bibfnamefont {G.~H.}\ \bibnamefont
		{Rawitscher}},\ }\href@noop {} {\bibfield  {journal} {\bibinfo  {journal}
		{Phys. Rev. C}\ }\textbf {\bibinfo {volume} {9}},\ \bibinfo {pages} {2210}
	(\bibinfo {year} {1974})}\BibitemShut {NoStop}%
\bibitem [{\citenamefont {Sakuragi}\ \emph {et~al.}(1986)\citenamefont
	{Sakuragi}, \citenamefont {Yahiro},\ and\ \citenamefont {Kamimura}}]{SYK86}%
\BibitemOpen
\bibfield  {author} {\bibinfo {author} {\bibfnamefont {Y.}~\bibnamefont
		{Sakuragi}}, \bibinfo {author} {\bibfnamefont {M.}~\bibnamefont {Yahiro}}, \
	and\ \bibinfo {author} {\bibfnamefont {M.}~\bibnamefont {Kamimura}},\
}\href@noop {} {\bibfield  {journal} {\bibinfo  {journal} {Prog. Theor. Phys.
		Suppl.}\ }\textbf {\bibinfo {volume} {89}},\ \bibinfo {pages} {136} (\bibinfo
{year} {1986})}\BibitemShut {NoStop}%
\bibitem [{\citenamefont {Austern}\ \emph {et~al.}(1987)\citenamefont
	{Austern}, \citenamefont {Iseri}, \citenamefont {Kamimura}, \citenamefont
	{Kawai}, \citenamefont {Rawitscher},\ and\ \citenamefont {Yahiro}}]{AIK87}%
\BibitemOpen
\bibfield  {author} {\bibinfo {author} {\bibfnamefont {N.}~\bibnamefont
		{Austern}}, \bibinfo {author} {\bibfnamefont {Y.}~\bibnamefont {Iseri}},
	\bibinfo {author} {\bibfnamefont {M.}~\bibnamefont {Kamimura}}, \bibinfo
	{author} {\bibfnamefont {M.}~\bibnamefont {Kawai}}, \bibinfo {author}
	{\bibfnamefont {G.}~\bibnamefont {Rawitscher}}, \ and\ \bibinfo {author}
	{\bibfnamefont {M.}~\bibnamefont {Yahiro}},\ }\href@noop {} {\bibfield
	{journal} {\bibinfo  {journal} {Phys. Rep.}\ }\textbf {\bibinfo {volume}
		{154}},\ \bibinfo {pages} {125} (\bibinfo {year} {1987})}\BibitemShut
{NoStop}%
\bibitem [{\citenamefont {Yahiro}\ \emph {et~al.}(2012)\citenamefont {Yahiro},
	\citenamefont {Ogata}, \citenamefont {Matsumoto},\ and\ \citenamefont
	{Minomo}}]{YOM12}%
\BibitemOpen
\bibfield  {author} {\bibinfo {author} {\bibfnamefont {M.}~\bibnamefont
		{Yahiro}}, \bibinfo {author} {\bibfnamefont {K.}~\bibnamefont {Ogata}},
	\bibinfo {author} {\bibfnamefont {T.}~\bibnamefont {Matsumoto}}, \ and\
	\bibinfo {author} {\bibfnamefont {K.}~\bibnamefont {Minomo}},\ }\href@noop {}
{\bibfield  {journal} {\bibinfo  {journal} {Prog. Theor. Exp. Phys.}\
	}\textbf {\bibinfo {volume} {2012}} (\bibinfo {year} {2012})}\BibitemShut
{NoStop}%
\bibitem [{\citenamefont {Matsumoto}\ \emph {et~al.}(2004)\citenamefont
	{Matsumoto}, \citenamefont {Hiyama}, \citenamefont {Ogata}, \citenamefont
	{Iseri}, \citenamefont {Kamimura}, \citenamefont {Chiba},\ and\ \citenamefont
	{Yahiro}}]{MHO04}%
\BibitemOpen
\bibfield  {author} {\bibinfo {author} {\bibfnamefont {T.}~\bibnamefont
		{Matsumoto}}, \bibinfo {author} {\bibfnamefont {E.}~\bibnamefont {Hiyama}},
	\bibinfo {author} {\bibfnamefont {K.}~\bibnamefont {Ogata}}, \bibinfo
	{author} {\bibfnamefont {Y.}~\bibnamefont {Iseri}}, \bibinfo {author}
	{\bibfnamefont {M.}~\bibnamefont {Kamimura}}, \bibinfo {author}
	{\bibfnamefont {S.}~\bibnamefont {Chiba}}, \ and\ \bibinfo {author}
	{\bibfnamefont {M.}~\bibnamefont {Yahiro}},\ }\href@noop {} {\bibfield
	{journal} {\bibinfo  {journal} {Phys. Rev. C}\ }\textbf {\bibinfo {volume}
		{70}},\ \bibinfo {pages} {061601} (\bibinfo {year} {2004})}\BibitemShut
{NoStop}%
\bibitem [{\citenamefont {Fern\'andez-Garc\'{\i}a}\ \emph
	{et~al.}(2015)\citenamefont {Fern\'andez-Garc\'{\i}a}, \citenamefont
	{Cubero}, \citenamefont {Acosta}, \citenamefont {Alcorta}, \citenamefont
	{Alvarez}, \citenamefont {Borge}, \citenamefont {Buchmann}, \citenamefont
	{Diget}, \citenamefont {Falou}, \citenamefont {Fulton}, \citenamefont
	{Fynbo}, \citenamefont {Galaviz}, \citenamefont {G\'omez-Camacho},
	\citenamefont {Kanungo}, \citenamefont {Lay}, \citenamefont {Madurga},
	\citenamefont {Martel}, \citenamefont {Moro}, \citenamefont {Mukha},
	\citenamefont {Nilsson}, \citenamefont {Rodr\'{\i}guez-Gallardo},
	\citenamefont {S\'anchez-Ben\'{\i}tez}, \citenamefont {Shotter},
	\citenamefont {Tengblad},\ and\ \citenamefont {Walden}}]{FCA15}%
\BibitemOpen
\bibfield  {author} {\bibinfo {author} {\bibfnamefont {J.~P.}\ \bibnamefont
		{Fern\'andez-Garc\'{\i}a}}, \bibinfo {author} {\bibfnamefont
		{M.}~\bibnamefont {Cubero}}, \bibinfo {author} {\bibfnamefont
		{L.}~\bibnamefont {Acosta}}, \bibinfo {author} {\bibfnamefont
		{M.}~\bibnamefont {Alcorta}}, \bibinfo {author} {\bibfnamefont {M.~A.~G.}\
		\bibnamefont {Alvarez}}, \bibinfo {author} {\bibfnamefont {M.~J.~G.}\
		\bibnamefont {Borge}}, \bibinfo {author} {\bibfnamefont {L.}~\bibnamefont
		{Buchmann}}, \bibinfo {author} {\bibfnamefont {C.~A.}\ \bibnamefont {Diget}},
	\bibinfo {author} {\bibfnamefont {H.~A.}\ \bibnamefont {Falou}}, \bibinfo
	{author} {\bibfnamefont {B.}~\bibnamefont {Fulton}}, \bibinfo {author}
	{\bibfnamefont {H.~O.~U.}\ \bibnamefont {Fynbo}}, \bibinfo {author}
	{\bibfnamefont {D.}~\bibnamefont {Galaviz}}, \bibinfo {author} {\bibfnamefont
		{J.}~\bibnamefont {G\'omez-Camacho}}, \bibinfo {author} {\bibfnamefont
		{R.}~\bibnamefont {Kanungo}}, \bibinfo {author} {\bibfnamefont {J.~A.}\
		\bibnamefont {Lay}}, \bibinfo {author} {\bibfnamefont {M.}~\bibnamefont
		{Madurga}}, \bibinfo {author} {\bibfnamefont {I.}~\bibnamefont {Martel}},
	\bibinfo {author} {\bibfnamefont {A.~M.}\ \bibnamefont {Moro}}, \bibinfo
	{author} {\bibfnamefont {I.}~\bibnamefont {Mukha}}, \bibinfo {author}
	{\bibfnamefont {T.}~\bibnamefont {Nilsson}}, \bibinfo {author} {\bibfnamefont
		{M.}~\bibnamefont {Rodr\'{\i}guez-Gallardo}}, \bibinfo {author}
	{\bibfnamefont {A.~M.}\ \bibnamefont {S\'anchez-Ben\'{\i}tez}}, \bibinfo
	{author} {\bibfnamefont {A.}~\bibnamefont {Shotter}}, \bibinfo {author}
	{\bibfnamefont {O.}~\bibnamefont {Tengblad}}, \ and\ \bibinfo {author}
	{\bibfnamefont {P.}~\bibnamefont {Walden}},\ }\href@noop {} {\bibfield
	{journal} {\bibinfo  {journal} {Phys. Rev. C}\ }\textbf {\bibinfo {volume}
		{92}},\ \bibinfo {pages} {044608} (\bibinfo {year} {2015})}\BibitemShut
{NoStop}%
\bibitem [{\citenamefont {Dasso}\ \emph {et~al.}(1996)\citenamefont {Dasso},
	\citenamefont {Lenzi},\ and\ \citenamefont {Vitturi}}]{DLV96}%
\BibitemOpen
\bibfield  {author} {\bibinfo {author} {\bibfnamefont {C.}~\bibnamefont
		{Dasso}}, \bibinfo {author} {\bibfnamefont {S.}~\bibnamefont {Lenzi}}, \ and\
	\bibinfo {author} {\bibfnamefont {A.}~\bibnamefont {Vitturi}},\ }\href
{\doibase http://dx.doi.org/10.1016/S0375-9474(96)00300-4} {\bibfield
	{journal} {\bibinfo  {journal} {Nucl. Phys. A}\ }\textbf {\bibinfo {volume}
		{611}},\ \bibinfo {pages} {124 } (\bibinfo {year} {1996})}\BibitemShut
{NoStop}%
\bibitem [{\citenamefont {Nunes}\ and\ \citenamefont {Thompson}(1999)}]{NT99}%
\BibitemOpen
\bibfield  {author} {\bibinfo {author} {\bibfnamefont {F.~M.}\ \bibnamefont
		{Nunes}}\ and\ \bibinfo {author} {\bibfnamefont {I.~J.}\ \bibnamefont
		{Thompson}},\ }\href@noop {} {\bibfield  {journal} {\bibinfo  {journal}
		{Phys. Rev. C}\ }\textbf {\bibinfo {volume} {59}},\ \bibinfo {pages} {2652}
	(\bibinfo {year} {1999})}\BibitemShut {NoStop}%
\bibitem [{\citenamefont {Tostevin}\ \emph {et~al.}(2001)\citenamefont
	{Tostevin}, \citenamefont {Nunes},\ and\ \citenamefont {Thompson}}]{TNT01}%
\BibitemOpen
\bibfield  {author} {\bibinfo {author} {\bibfnamefont {J.~A.}\ \bibnamefont
		{Tostevin}}, \bibinfo {author} {\bibfnamefont {F.~M.}\ \bibnamefont {Nunes}},
	\ and\ \bibinfo {author} {\bibfnamefont {I.~J.}\ \bibnamefont {Thompson}},\
}\href@noop {} {\bibfield  {journal} {\bibinfo  {journal} {Phys. Rev. C}\
}\textbf {\bibinfo {volume} {63}},\ \bibinfo {pages} {024617} (\bibinfo
{year} {2001})}\BibitemShut {NoStop}%
\bibitem [{\citenamefont {Goldstein}\ \emph {et~al.}(2006)\citenamefont
	{Goldstein}, \citenamefont {Baye},\ and\ \citenamefont {Capel}}]{GBC06}%
\BibitemOpen
\bibfield  {author} {\bibinfo {author} {\bibfnamefont {G.}~\bibnamefont
		{Goldstein}}, \bibinfo {author} {\bibfnamefont {D.}~\bibnamefont {Baye}}, \
	and\ \bibinfo {author} {\bibfnamefont {P.}~\bibnamefont {Capel}},\
}\href@noop {} {\bibfield  {journal} {\bibinfo  {journal} {Phys. Rev. C}\
}\textbf {\bibinfo {volume} {73}},\ \bibinfo {pages} {024602} (\bibinfo
{year} {2006})}\BibitemShut {NoStop}%
\bibitem [{\citenamefont {Marta}\ \emph {et~al.}(2008)\citenamefont {Marta},
	\citenamefont {Canto},\ and\ \citenamefont {Donangelo}}]{MCD08}%
\BibitemOpen
\bibfield  {author} {\bibinfo {author} {\bibfnamefont {H.~D.}\ \bibnamefont
		{Marta}}, \bibinfo {author} {\bibfnamefont {L.~F.}\ \bibnamefont {Canto}}, \
	and\ \bibinfo {author} {\bibfnamefont {R.}~\bibnamefont {Donangelo}},\ }\href
{\doibase 10.1103/PhysRevC.78.034612} {\bibfield  {journal} {\bibinfo
		{journal} {Phys. Rev. C}\ }\textbf {\bibinfo {volume} {78}},\ \bibinfo
	{pages} {034612} (\bibinfo {year} {2008})}\BibitemShut {NoStop}%
\bibitem [{\citenamefont {Crespo}\ \emph {et~al.}(2011)\citenamefont {Crespo},
	\citenamefont {Deltuva},\ and\ \citenamefont {Moro}}]{CDM11}%
\BibitemOpen
\bibfield  {author} {\bibinfo {author} {\bibfnamefont {R.}~\bibnamefont
		{Crespo}}, \bibinfo {author} {\bibfnamefont {A.}~\bibnamefont {Deltuva}}, \
	and\ \bibinfo {author} {\bibfnamefont {A.~M.}\ \bibnamefont {Moro}},\ }\href
{\doibase 10.1103/PhysRevC.83.044622} {\bibfield  {journal} {\bibinfo
		{journal} {Phys. Rev. C}\ }\textbf {\bibinfo {volume} {83}},\ \bibinfo
	{pages} {044622} (\bibinfo {year} {2011})}\BibitemShut {NoStop}%
\bibitem [{\citenamefont {Kucuk}\ and\ \citenamefont {Moro}(2012)}]{KM12}%
\BibitemOpen
\bibfield  {author} {\bibinfo {author} {\bibfnamefont {Y.}~\bibnamefont
		{Kucuk}}\ and\ \bibinfo {author} {\bibfnamefont {A.~M.}\ \bibnamefont
		{Moro}},\ }\href@noop {} {\bibfield  {journal} {\bibinfo  {journal} {Phys.
			Rev. C}\ }\textbf {\bibinfo {volume} {86}},\ \bibinfo {pages} {034601}
	(\bibinfo {year} {2012})}\BibitemShut {NoStop}%
\bibitem [{\citenamefont {Rangel}\ \emph {et~al.}(2016)\citenamefont {Rangel},
	\citenamefont {Lubian}, \citenamefont {Canto},\ and\ \citenamefont
	{Gomes}}]{RLC16}%
\BibitemOpen
\bibfield  {author} {\bibinfo {author} {\bibfnamefont {J.}~\bibnamefont
		{Rangel}}, \bibinfo {author} {\bibfnamefont {J.}~\bibnamefont {Lubian}},
	\bibinfo {author} {\bibfnamefont {L.~F.}\ \bibnamefont {Canto}}, \ and\
	\bibinfo {author} {\bibfnamefont {P.~R.~S.}\ \bibnamefont {Gomes}},\ }\href
{\doibase 10.1103/PhysRevC.93.054610} {\bibfield  {journal} {\bibinfo
		{journal} {Phys. Rev. C}\ }\textbf {\bibinfo {volume} {93}},\ \bibinfo
	{pages} {054610} (\bibinfo {year} {2016})}\BibitemShut {NoStop}%
\bibitem [{\citenamefont {Dasso}\ \emph {et~al.}(1999)\citenamefont {Dasso},
	\citenamefont {Lenzi},\ and\ \citenamefont {Vitturi}}]{DLV99}%
\BibitemOpen
\bibfield  {author} {\bibinfo {author} {\bibfnamefont {C.~H.}\ \bibnamefont
		{Dasso}}, \bibinfo {author} {\bibfnamefont {S.~M.}\ \bibnamefont {Lenzi}}, \
	and\ \bibinfo {author} {\bibfnamefont {A.}~\bibnamefont {Vitturi}},\ }\href
{\doibase 10.1103/PhysRevC.59.539} {\bibfield  {journal} {\bibinfo  {journal}
		{Phys. Rev. C}\ }\textbf {\bibinfo {volume} {59}},\ \bibinfo {pages} {539}
	(\bibinfo {year} {1999})}\BibitemShut {NoStop}%
\bibitem [{\citenamefont {Hussein}\ \emph {et~al.}(2006)\citenamefont
	{Hussein}, \citenamefont {Lichtenth\"aler}, \citenamefont {Nunes},\ and\
	\citenamefont {Thompson}}]{HLN06}%
\BibitemOpen
\bibfield  {author} {\bibinfo {author} {\bibfnamefont {M.~S.}\ \bibnamefont
		{Hussein}}, \bibinfo {author} {\bibfnamefont {R.}~\bibnamefont
		{Lichtenth\"aler}}, \bibinfo {author} {\bibfnamefont {F.~M.}\ \bibnamefont
		{Nunes}}, \ and\ \bibinfo {author} {\bibfnamefont {I.~J.}\ \bibnamefont
		{Thompson}},\ }\href {\doibase
	http://dx.doi.org/10.1016/j.physletb.2006.07.046} {\bibfield  {journal}
	{\bibinfo  {journal} {Phys. Lett. B}\ }\textbf {\bibinfo {volume} {640}},\
	\bibinfo {pages} {91 } (\bibinfo {year} {2006})}\BibitemShut {NoStop}%
\bibitem [{\citenamefont {Kumar}\ and\ \citenamefont
	{Bonaccorso}(2012)}]{KB12}%
\BibitemOpen
\bibfield  {author} {\bibinfo {author} {\bibfnamefont {R.}~\bibnamefont
		{Kumar}}\ and\ \bibinfo {author} {\bibfnamefont {A.}~\bibnamefont
		{Bonaccorso}},\ }\href {\doibase 10.1103/PhysRevC.86.061601} {\bibfield
	{journal} {\bibinfo  {journal} {Phys. Rev. C}\ }\textbf {\bibinfo {volume}
		{86}},\ \bibinfo {pages} {061601} (\bibinfo {year} {2012})}\BibitemShut
{NoStop}%
\bibitem [{\citenamefont {Dasso}\ \emph {et~al.}(1998)\citenamefont {Dasso},
	\citenamefont {Lenzi},\ and\ \citenamefont {Vitturi}}]{DLV98}%
\BibitemOpen
\bibfield  {author} {\bibinfo {author} {\bibfnamefont {C.}~\bibnamefont
		{Dasso}}, \bibinfo {author} {\bibfnamefont {S.}~\bibnamefont {Lenzi}}, \ and\
	\bibinfo {author} {\bibfnamefont {A.}~\bibnamefont {Vitturi}},\ }\href
{\doibase http://dx.doi.org/10.1016/S0375-9474(98)00420-5} {\bibfield
	{journal} {\bibinfo  {journal} {Nucl. Phys. A}\ }\textbf {\bibinfo {volume}
		{639}},\ \bibinfo {pages} {635 } (\bibinfo {year} {1998})}\BibitemShut
{NoStop}%
\bibitem [{\citenamefont {Canto}\ \emph {et~al.}(2015)\citenamefont {Canto},
	\citenamefont {Gomes}, \citenamefont {Donangelo}, \citenamefont {Lubian},\
	and\ \citenamefont {Hussein}}]{CGD15}%
\BibitemOpen
\bibfield  {author} {\bibinfo {author} {\bibfnamefont {L.~F.}\ \bibnamefont
		{Canto}}, \bibinfo {author} {\bibfnamefont {P.~R.~S.}\ \bibnamefont {Gomes}},
	\bibinfo {author} {\bibfnamefont {R.}~\bibnamefont {Donangelo}}, \bibinfo
	{author} {\bibfnamefont {J.}~\bibnamefont {Lubian}}, \ and\ \bibinfo {author}
	{\bibfnamefont {M.~S.}\ \bibnamefont {Hussein}},\ }\href {\doibase
	http://dx.doi.org/10.1016/j.physrep.2015.08.001} {\bibfield  {journal}
	{\bibinfo  {journal} {Phys. Rep.}\ }\textbf {\bibinfo {volume} {596}},\
	\bibinfo {pages} {1} (\bibinfo {year} {2015})}\BibitemShut {NoStop}%
\bibitem [{\citenamefont {Descouvemont}\ and\ \citenamefont
	{Dufour}(2012)}]{DD12}%
\BibitemOpen
\bibfield  {author} {\bibinfo {author} {\bibfnamefont {P.}~\bibnamefont
		{Descouvemont}}\ and\ \bibinfo {author} {\bibfnamefont {M.}~\bibnamefont
		{Dufour}},\ }\href@noop {} {\emph {\bibinfo {title} {Clusters in Nuclei}}},\
edited by\ \bibinfo {editor} {\bibfnamefont {C.}~\bibnamefont {Beck}},\
Vol.~\bibinfo {volume} {2}\ (\bibinfo  {publisher} {Springer},\ \bibinfo
{year} {2012})\BibitemShut {NoStop}%
\bibitem [{\citenamefont {Baye}(2015)}]{Ba15}%
\BibitemOpen
\bibfield  {author} {\bibinfo {author} {\bibfnamefont {D.}~\bibnamefont
		{Baye}},\ }\href {\doibase http://dx.doi.org/10.1016/j.physrep.2014.11.006}
{\bibfield  {journal} {\bibinfo  {journal} {Phys. Rep.}\ }\textbf {\bibinfo
		{volume} {565}},\ \bibinfo {pages} {1} (\bibinfo {year} {2015})}\BibitemShut
{NoStop}%
\bibitem [{\citenamefont {Hiyama}\ \emph {et~al.}(2003)\citenamefont {Hiyama},
	\citenamefont {Kino},\ and\ \citenamefont {Kamimura}}]{HKK03}%
\BibitemOpen
\bibfield  {author} {\bibinfo {author} {\bibfnamefont {E.}~\bibnamefont
		{Hiyama}}, \bibinfo {author} {\bibfnamefont {Y.}~\bibnamefont {Kino}}, \ and\
	\bibinfo {author} {\bibfnamefont {M.}~\bibnamefont {Kamimura}},\ }\href
{\doibase http://dx.doi.org/10.1016/S0146-6410(03)90015-9} {\bibfield
	{journal} {\bibinfo  {journal} {Prog. Part. Nucl. Phys.}\ }\textbf {\bibinfo
		{volume} {51}},\ \bibinfo {pages} {223 } (\bibinfo {year}
	{2003})}\BibitemShut {NoStop}%
\bibitem [{\citenamefont {Thompson}(1988)}]{Th88}%
\BibitemOpen
\bibfield  {author} {\bibinfo {author} {\bibfnamefont {I.~J.}\ \bibnamefont
		{Thompson}},\ }\href@noop {} {\bibfield  {journal} {\bibinfo  {journal}
		{Comput. Phys. Rep.}\ }\textbf {\bibinfo {volume} {7}},\ \bibinfo {pages}
	{167} (\bibinfo {year} {1988})}\BibitemShut {NoStop}%
\bibitem [{\citenamefont {Thompson}(2010)}]{Th10}%
\BibitemOpen
\bibfield  {author} {\bibinfo {author} {\bibfnamefont {I.~J.}\ \bibnamefont
		{Thompson}},\ }\href@noop {} {\emph {\bibinfo {title} {NIST Handbook of
			Mathematical Functions}}},\ edited by\ \bibinfo {editor} {\bibfnamefont
	{F.~W.~J.}\ \bibnamefont {Olver}}, \bibinfo {editor} {\bibfnamefont {D.~W.}\
	\bibnamefont {Lozier}}, \bibinfo {editor} {\bibfnamefont {R.~F.}\
	\bibnamefont {Boisvert}}, \ and\ \bibinfo {editor} {\bibfnamefont {C.~W.}\
	\bibnamefont {Clark}}\ (\bibinfo  {publisher} {Cambridge University Press},\
\bibinfo {year} {2010})\ p.\ \bibinfo {pages} {741}\BibitemShut {NoStop}%
\bibitem [{\citenamefont {Descouvemont}\ and\ \citenamefont
	{Baye}(2010)}]{DB10}%
\BibitemOpen
\bibfield  {author} {\bibinfo {author} {\bibfnamefont {P.}~\bibnamefont
		{Descouvemont}}\ and\ \bibinfo {author} {\bibfnamefont {D.}~\bibnamefont
		{Baye}},\ }\href@noop {} {\bibfield  {journal} {\bibinfo  {journal} {Rep.
			Prog. Phys.}\ }\textbf {\bibinfo {volume} {73}},\ \bibinfo {pages} {036301}
	(\bibinfo {year} {2010})}\BibitemShut {NoStop}%
\bibitem [{\citenamefont {Descouvemont}(2016)}]{De16a}%
\BibitemOpen
\bibfield  {author} {\bibinfo {author} {\bibfnamefont {P.}~\bibnamefont
		{Descouvemont}},\ }\href {\doibase
	http://dx.doi.org/10.1016/j.cpc.2015.10.015} {\bibfield  {journal} {\bibinfo
		{journal} {Comput. Phys. Commun.}\ }\textbf {\bibinfo {volume} {200}},\
	\bibinfo {pages} {199} (\bibinfo {year} {2016})}\BibitemShut {NoStop}%
\bibitem [{\citenamefont {Canto}\ and\ \citenamefont {Hussein}(2013)}]{CH13}%
\BibitemOpen
\bibfield  {author} {\bibinfo {author} {\bibfnamefont {L.~F.}\ \bibnamefont
		{Canto}}\ and\ \bibinfo {author} {\bibfnamefont {M.~S.}\ \bibnamefont
		{Hussein}},\ }\href@noop {} {\emph {\bibinfo {title} {Scattering Theory of
			Molecules, Atoms and Nuclei}}}\ (\bibinfo  {publisher} {World Scientific
	Publishing, Singapore},\ \bibinfo {year} {2013})\BibitemShut {NoStop}%
\bibitem [{\citenamefont {Druet}\ and\ \citenamefont
	{Descouvemont}(2012)}]{DD12b}%
\BibitemOpen
\bibfield  {author} {\bibinfo {author} {\bibfnamefont {T.}~\bibnamefont
		{Druet}}\ and\ \bibinfo {author} {\bibfnamefont {P.}~\bibnamefont
		{Descouvemont}},\ }\href@noop {} {\bibfield  {journal} {\bibinfo  {journal}
		{Eur. Phys. J. A}\ }\textbf {\bibinfo {volume} {48}},\ \bibinfo {pages} {147}
	(\bibinfo {year} {2012})}\BibitemShut {NoStop}%
\bibitem [{\citenamefont {Hussein}\ and\ \citenamefont {Marques}(1986)}]{HM86}%
\BibitemOpen
\bibfield  {author} {\bibinfo {author} {\bibfnamefont {M.}~\bibnamefont
		{Hussein}}\ and\ \bibinfo {author} {\bibfnamefont {G.}~\bibnamefont
		{Marques}},\ }\href {\doibase http://dx.doi.org/10.1016/0003-4916(86)90024-2}
{\bibfield  {journal} {\bibinfo  {journal} {Ann. Phys.}\ }\textbf {\bibinfo
		{volume} {172}},\ \bibinfo {pages} {191 } (\bibinfo {year}
	{1986})}\BibitemShut {NoStop}%
\bibitem [{\citenamefont {Rhoades-Brown}\ \emph {et~al.}(1980)\citenamefont
	{Rhoades-Brown}, \citenamefont {Macfarlane},\ and\ \citenamefont
	{Pieper}}]{RMP80b}%
\BibitemOpen
\bibfield  {author} {\bibinfo {author} {\bibfnamefont {M.}~\bibnamefont
		{Rhoades-Brown}}, \bibinfo {author} {\bibfnamefont {M.~H.}\ \bibnamefont
		{Macfarlane}}, \ and\ \bibinfo {author} {\bibfnamefont {S.~C.}\ \bibnamefont
		{Pieper}},\ }\href@noop {} {\bibfield  {journal} {\bibinfo  {journal} {Phys.
			Rev. C}\ }\textbf {\bibinfo {volume} {21}},\ \bibinfo {pages} {2436}
	(\bibinfo {year} {1980})}\BibitemShut {NoStop}%
\bibitem [{\citenamefont {Thompson}\ and\ \citenamefont {Nunes}(2009)}]{TN09}%
\BibitemOpen
\bibfield  {author} {\bibinfo {author} {\bibfnamefont {I.}~\bibnamefont
		{Thompson}}\ and\ \bibinfo {author} {\bibfnamefont {F.}~\bibnamefont
		{Nunes}},\ }\href@noop {} {\emph {\bibinfo {title} {Nuclear Reactions for
			Astrophysics: Principles, Calculation and Applications of Low-Energy
			Reactions}}}\ (\bibinfo  {publisher} {Cambridge University Press},\ \bibinfo
{year} {2009})\BibitemShut {NoStop}%
\bibitem [{\citenamefont {Alder}\ and\ \citenamefont {Winther}(1975)}]{AW75}%
\BibitemOpen
\bibfield  {author} {\bibinfo {author} {\bibfnamefont {K.}~\bibnamefont
		{Alder}}\ and\ \bibinfo {author} {\bibfnamefont {A.}~\bibnamefont
		{Winther}},\ }\href@noop {} {\emph {\bibinfo {title} {Electromagnetic
			excitation : theory of Coulomb excitation with heavy ions}}}\ (\bibinfo
{publisher} {North-Holland, Amsterdam},\ \bibinfo {year} {1975})\BibitemShut
{NoStop}%
\bibitem [{\citenamefont {Alder}\ \emph {et~al.}(1956)\citenamefont {Alder},
	\citenamefont {Bohr}, \citenamefont {Huus}, \citenamefont {Mottelson},\ and\
	\citenamefont {Winther}}]{ABH56}%
\BibitemOpen
\bibfield  {author} {\bibinfo {author} {\bibfnamefont {K.}~\bibnamefont
		{Alder}}, \bibinfo {author} {\bibfnamefont {A.}~\bibnamefont {Bohr}},
	\bibinfo {author} {\bibfnamefont {T.}~\bibnamefont {Huus}}, \bibinfo {author}
	{\bibfnamefont {B.}~\bibnamefont {Mottelson}}, \ and\ \bibinfo {author}
	{\bibfnamefont {A.}~\bibnamefont {Winther}},\ }\href@noop {} {\bibfield
	{journal} {\bibinfo  {journal} {Rev. Mod. Phys.}\ }\textbf {\bibinfo {volume}
		{28}},\ \bibinfo {pages} {432} (\bibinfo {year} {1956})}\BibitemShut
{NoStop}%
\bibitem [{\citenamefont {Ohsaki}\ \emph {et~al.}(2002)\citenamefont {Ohsaki},
	\citenamefont {Igarashi}, \citenamefont {Kai},\ and\ \citenamefont
	{Nakazaki}}]{OIK02}%
\BibitemOpen
\bibfield  {author} {\bibinfo {author} {\bibfnamefont {A.}~\bibnamefont
		{Ohsaki}}, \bibinfo {author} {\bibfnamefont {A.}~\bibnamefont {Igarashi}},
	\bibinfo {author} {\bibfnamefont {T.}~\bibnamefont {Kai}}, \ and\ \bibinfo
	{author} {\bibfnamefont {S.}~\bibnamefont {Nakazaki}},\ }\href {\doibase
	http://dx.doi.org/10.1016/S0010-4655(02)00458-7} {\bibfield  {journal}
	{\bibinfo  {journal} {Comp. Phys. Commun.}\ }\textbf {\bibinfo {volume}
		{147}},\ \bibinfo {pages} {826 } (\bibinfo {year} {2002})}\BibitemShut
{NoStop}%
\bibitem [{\citenamefont {Buck}\ and\ \citenamefont {Merchant}(1988)}]{BM88}%
\BibitemOpen
\bibfield  {author} {\bibinfo {author} {\bibfnamefont {B.}~\bibnamefont
		{Buck}}\ and\ \bibinfo {author} {\bibfnamefont {A.~C.}\ \bibnamefont
		{Merchant}},\ }\href@noop {} {\bibfield  {journal} {\bibinfo  {journal} {J.
			Phys. G}\ }\textbf {\bibinfo {volume} {14}},\ \bibinfo {pages} {L211}
	(\bibinfo {year} {1988})}\BibitemShut {NoStop}%
\bibitem [{\citenamefont {Horiuchi}(1977)}]{Ho77}%
\BibitemOpen
\bibfield  {author} {\bibinfo {author} {\bibfnamefont {H.}~\bibnamefont
		{Horiuchi}},\ }\href@noop {} {\bibfield  {journal} {\bibinfo  {journal}
		{Prog. Theor. Phys. Suppl.}\ }\textbf {\bibinfo {volume} {62}},\ \bibinfo
	{pages} {90} (\bibinfo {year} {1977})}\BibitemShut {NoStop}%
\bibitem [{\citenamefont {Avrigeanu}\ \emph {et~al.}(2009)\citenamefont
	{Avrigeanu}, \citenamefont {Obreja}, \citenamefont {Roman}, \citenamefont
	{Avrigeanu},\ and\ \citenamefont {von Oertzen}}]{AOR09}%
\BibitemOpen
\bibfield  {author} {\bibinfo {author} {\bibfnamefont {M.}~\bibnamefont
		{Avrigeanu}}, \bibinfo {author} {\bibfnamefont {A.}~\bibnamefont {Obreja}},
	\bibinfo {author} {\bibfnamefont {F.}~\bibnamefont {Roman}}, \bibinfo
	{author} {\bibfnamefont {V.}~\bibnamefont {Avrigeanu}}, \ and\ \bibinfo
	{author} {\bibfnamefont {W.}~\bibnamefont {von Oertzen}},\ }\href {\doibase
	http://dx.doi.org/10.1016/j.adt.2009.02.001} {\bibfield  {journal} {\bibinfo
		{journal} {At. Data Nucl. Data Tables}\ }\textbf {\bibinfo {volume} {95}},\
	\bibinfo {pages} {501 } (\bibinfo {year} {2009})}\BibitemShut {NoStop}%
\bibitem [{\citenamefont {Pang}\ \emph {et~al.}(2009)\citenamefont {Pang},
	\citenamefont {Roussel-Chomaz}, \citenamefont {Savajols}, \citenamefont
	{Varner},\ and\ \citenamefont {Wolski}}]{PRS09}%
\BibitemOpen
\bibfield  {author} {\bibinfo {author} {\bibfnamefont {D.~Y.}\ \bibnamefont
		{Pang}}, \bibinfo {author} {\bibfnamefont {P.}~\bibnamefont
		{Roussel-Chomaz}}, \bibinfo {author} {\bibfnamefont {H.}~\bibnamefont
		{Savajols}}, \bibinfo {author} {\bibfnamefont {R.~L.}\ \bibnamefont
		{Varner}}, \ and\ \bibinfo {author} {\bibfnamefont {R.}~\bibnamefont
		{Wolski}},\ }\href {\doibase 10.1103/PhysRevC.79.024615} {\bibfield
	{journal} {\bibinfo  {journal} {Phys. Rev. C}\ }\textbf {\bibinfo {volume}
		{79}},\ \bibinfo {pages} {024615} (\bibinfo {year} {2009})}\BibitemShut
{NoStop}%
\bibitem [{\citenamefont {Frahn}(1966)}]{Fr66}%
\BibitemOpen
\bibfield  {author} {\bibinfo {author} {\bibfnamefont {W.}~\bibnamefont
		{Frahn}},\ }\href@noop {} {\bibfield  {journal} {\bibinfo  {journal} {Nucl.
			Phys.}\ }\textbf {\bibinfo {volume} {75}},\ \bibinfo {pages} {577 } (\bibinfo
	{year} {1966})}\BibitemShut {NoStop}%
\bibitem [{\citenamefont {Frahn}(1985)}]{Fr85}%
\BibitemOpen
\bibfield  {author} {\bibinfo {author} {\bibfnamefont {W.~E.}\ \bibnamefont
		{Frahn}},\ }\href@noop {} {\emph {\bibinfo {title} {Diffractive processes in
			nuclear physics}}}\ (\bibinfo  {publisher} {Clarendon press, Oxford},\
\bibinfo {year} {1985})\BibitemShut {NoStop}%
\bibitem [{\citenamefont {Camacho}\ \emph {et~al.}(2016)\citenamefont
	{Camacho}, \citenamefont {Diaz-Torres}, \citenamefont {Gomes},\ and\
	\citenamefont {Lubian}}]{CDG16}%
\BibitemOpen
\bibfield  {author} {\bibinfo {author} {\bibfnamefont {A.~G.}\ \bibnamefont
		{Camacho}}, \bibinfo {author} {\bibfnamefont {A.}~\bibnamefont
		{Diaz-Torres}}, \bibinfo {author} {\bibfnamefont {P.~R.~S.}\ \bibnamefont
		{Gomes}}, \ and\ \bibinfo {author} {\bibfnamefont {J.}~\bibnamefont
		{Lubian}},\ }\href {\doibase 10.1103/PhysRevC.93.024604} {\bibfield
	{journal} {\bibinfo  {journal} {Phys. Rev. C}\ }\textbf {\bibinfo {volume}
		{93}},\ \bibinfo {pages} {024604} (\bibinfo {year} {2016})}\BibitemShut
{NoStop}%
\bibitem [{\citenamefont {Sopkovich}(1962)}]{So62}%
\BibitemOpen
\bibfield  {author} {\bibinfo {author} {\bibfnamefont {M.~J.}\ \bibnamefont
		{Sopkovich}},\ }\href@noop {} {\bibfield  {journal} {\bibinfo  {journal}
		{Nuovo Cimento}\ }\textbf {\bibinfo {volume} {26}},\ \bibinfo {pages} {186}
	(\bibinfo {year} {1962})}\BibitemShut {NoStop}%
\bibitem [{\citenamefont {Baye}\ \emph {et~al.}(1991)\citenamefont {Baye},
	\citenamefont {Sauwens}, \citenamefont {Descouvemont},\ and\ \citenamefont
	{Keller}}]{BSD91}%
\BibitemOpen
\bibfield  {author} {\bibinfo {author} {\bibfnamefont {D.}~\bibnamefont
		{Baye}}, \bibinfo {author} {\bibfnamefont {C.}~\bibnamefont {Sauwens}},
	\bibinfo {author} {\bibfnamefont {P.}~\bibnamefont {Descouvemont}}, \ and\
	\bibinfo {author} {\bibfnamefont {S.}~\bibnamefont {Keller}},\ }\href@noop {}
{\bibfield  {journal} {\bibinfo  {journal} {Nucl. Phys. A}\ }\textbf
	{\bibinfo {volume} {529}},\ \bibinfo {pages} {467} (\bibinfo {year}
	{1991})}\BibitemShut {NoStop}%
\end{thebibliography}
%

\end{document}